\documentclass[usenatbib,useAMS,letterpaper]{mn2e}
\usepackage{graphicx}
\usepackage{epstopdf}
\usepackage{ctable}
\usepackage{url}
\usepackage{times}
\usepackage{xspace}
\usepackage[T1]{fontenc}
\usepackage{aecompl}

\newcommand{\kmpers}{~\mathrm{km ~s}^{-1}}

\newcommand{\msun}{~\mathrm{M_{\odot}}}

\newcommand{\sfr}{\dot{M}_{\star}}
\newcommand{\mwind}{\dot{M}_{\mathrm{out}}}

\newcommand{\mgas}{M_{\rm g}}
\newcommand{\mstar}{M_{\star}}

\newcommand{\vesc}{v_{\mathrm{esc}}}

\newcommand{\sigsfr}{\dot{\Sigma}_{\star}}
\newcommand{\sigsfrturb}{\dot{\Sigma}_{\star}^{\mathrm{turb}}}
\newcommand{\sigsfrtherm}{\dot{\Sigma}_{\star}^{\mathrm{th}}}
\newcommand{\sigpdot}{\dot{\Sigma}_{P}}
\newcommand{\siggas}{\Sigma_{\mathrm{g}}}
\newcommand{\msiggas}{\langle\siggas\rangle}

\newcommand{\siggasmax}{\Sigma_{\mathrm{g}}^{\mathrm{max}}}
\newcommand{\siggasmaxturb}{\Sigma_{\mathrm{g, turb}}^{\mathrm{max}}}
\newcommand{\siggasmaxtherm}{\Sigma_{\mathrm{g, th}}^{\mathrm{max}}}
\newcommand{\sigwind}{\dot{\Sigma}_{\mathrm{out}}}
\newcommand{\sigt}{\sigma_{\mathrm{T}}}
\newcommand{\sigss}{\Sigma_{\mathrm{self-shield}}}

\newcommand{\cs}{c_{\mathrm{s}}}
\newcommand{\vc}{v_{\mathrm{c,gal}}}
\newcommand{\vchalo}{v_{\mathrm{c}}}
\newcommand{\fw}{f_{\mathrm{out}}}
\newcommand{\xw}{x_{\mathrm{out}}}
\newcommand{\fwtherm}{f_{\mathrm{out}}^{\mathrm{th}}}
\newcommand{\xwtherm}{x_{\mathrm{out}}^{\mathrm{th}}}
\newcommand{\fwturb}{f_{\mathrm{out}}^{\mathrm{turb}}}
\newcommand{\xwturb}{x_{\mathrm{out}}^{\mathrm{turb}}}
\newcommand{\fg}{f_{\mathrm{g}}}
\newcommand{\re}{R_{\mathrm{e}}}

\newcommand{\qturb}{Q_{\mathrm{turb}}}
\newcommand{\qtherm}{Q_{\mathrm{th}}}
\newcommand{\etatherm}{\eta_{\mathrm{th}}}
\newcommand{\etaturb}{\eta_{\mathrm{turb}}}
\newcommand{\mtherm}{\mathcal{M}_{\mathrm{th}}}
\newcommand{\mturb}{\mathcal{M}_{\mathrm{turb}}}

\newcommand{\acknowledgments}{\begin{small}\section*{Acknowledgements}\end{small}}

\newcommand\sref[1]{\hyperref[#1]{Section~\ref*{#1}}}
\newcommand\fref[1]{\hyperref[#1]{Fig.~\ref*{#1}}}
\newcommand\aref[1]{\hyperref[#1]{Appendix~\ref*{#1}}}

\newcommand\rev[1]{#1}

\title[Self-regulated star formation \& outflows]{How stellar feedback simultaneously regulates star formation and drives outflows}
\author[C.~C. Hayward \& P.~F. Hopkins]{
\parbox[t]{\textwidth}{
Christopher C. Hayward$^{1,2,3}$\thanks{E-mail: chayward@simonsfoundation.org} \&
Philip F. Hopkins$^2$
}
\vspace*{6pt} \\
$^1$Center for Computational Astrophysics, Flatiron Institute, 162 Fifth Avenue, New York, NY 10010, USA\\
$^2$TAPIR 350-17, California Institute of Technology, 1200 E. California Boulevard, Pasadena, CA 91125, USA\\
$^3$Harvard--Smithsonian Center for Astrophysics, 60 Garden Street, Cambridge, MA 02138, USA}

\usepackage[breaklinks,colorlinks,allcolors=blue]{hyperref}
\usepackage[all]{hypcap}

\begin{document}

\date{Submitted to MNRAS}

\pagerange{\pageref{firstpage}--\pageref{lastpage}} \pubyear{2016}

\maketitle

\label{firstpage}

\begin{abstract}
We present an analytic model for how momentum deposition from stellar feedback simultaneously regulates star formation and drives outflows
in a turbulent interstellar medium (ISM). Because the ISM is turbulent, a given patch of ISM
exhibits sub-patches with a range of surface densities. The high-density patches are `pushed' by feedback, thereby driving turbulence and
self-regulating local star formation. Sufficiently low-density patches, however, are accelerated to above the escape velocity before the region
can self-adjust and are thus vented as outflows.
When the gas fraction is $\ga 0.3$, the ratio of the turbulent velocity dispersion to the circular
velocity is sufficiently high that at any given time, of order half of the ISM has surface density less than the critical value and thus can be blown out on
a dynamical time. The resulting outflows have a mass-loading factor ($\eta \equiv \mwind/\mstar$) that is inversely proportional to the gas fraction
times the circular velocity.
At low gas fractions, the star formation rate needed for local self-regulation, and corresponding turbulent Mach number, decline rapidly;
the ISM is `smoother', and it is actually \emph{more} difficult to drive winds with large mass-loading factors.
Crucially, our model predicts that stellar-feedback-driven outflows should be suppressed at $z \la 1$ in $\mstar \ga 10^{10} \msun$ galaxies.
This mechanism allows massive galaxies to exhibit violent outflows at high redshifts and then `shut down' those outflows at late times,
thereby enabling the formation of a smooth, extended thin stellar disk.
We provide simple fitting functions for $\eta$ that should be useful for sub-resolution and semi-analytic models.
\end{abstract}


\section{Introduction} \label{S:intro}

It is widely believed that stellar feedback is crucial for regulating star formation. In the absence
of feedback, gas would be completely converted to stars on a free-fall time. In contrast, the actual
efficiency is only a few per cent (e.g. \citealt{Bigiel2008,Bigiel2011,Krumholz2012,Leroy2013}),
although the efficiency per free-fall time can be as high as unity for individual giant molecular clouds (GMCs; \citealt{Murray2011}).
Similarly, without feedback, one would expect most gas in dark matter halos to have been converted
into stars by the present day.
Thus, the ratio of stellar mass to halo mass should be equal to the
baryon fraction \citep[e.g.][]{Cole1991,WhiteFrenk1991,Blanchard1992,Balogh2001};
however, for all halo masses, the stellar-to-halo mass ratio is much less than
this value \citep[e.g.][]{Conroy2009,Moster2013,Behroozi2013}
To avoid over-producing stars in numerical simulations and semi-analytic models (SAMs) of
galaxy formation, it is necessary to invoke some form(s) of strong stellar feedback (see
\citealt{Somerville2014} for a recent review). It has long been recognised that supernova feedback
can generate strong outflows and thus
may provide a solution to this overcooling problem \citep[e.g.][]{MathewsBaker1971,Larson1974,DekelSilk1986,WhiteRees1978,WhiteFrenk1991}.
Other potentially important stellar feedback channels include
stellar winds, radiation pressure, and photoionization (see \citealt{Hopkins2012ISM} and references therein), \rev{in addition to cosmic
rays \citep[e.g.][]{Uhlig2012,Hanasz2013,Booth2013,Salem2014a,Salem2014b,Girichidis2016}}.
This feedback stabilizes the interstellar medium (ISM) such that only a few per cent of the gas mass is converted into stars per orbital time and
the Kennicutt-Schmidt relation \citep{Schmidt1959,Kennicutt1998} can be reproduced
\citep[e.g.][]{Stinson2006,ShettyOstriker2008,ShettyOstriker2012,KKO2011,OstrikerShetty2011,Hopkins2012ISM,Hopkins2014,KOK2013,
Agertz2015a}. Moreover, feedback is thought to power strong outflows that prevent most of the gas accreted
onto galaxy halos from ever forming stars: for most galaxy masses, the required `mass-loading factor', $\eta$,
which is the ratio of the mass outflow rate to the star formation rate (SFR), must be significantly greater than unity
\citep[e.g.][]{Benson2014,Lu2014,Lu2015,Mitra2015}.

There is an increasing amount of observational evidence for the existence of stellar-feedback-driven
outflows (e.g.~\citealt{LyndsSandage1963,Burbidge1964,Heckman1990,Heckman2000,Heckman2015,
Martin1996,Martin1998,Martin1999,Martin2005,Martin2006,Weiner2009,Steidel2010,Bordoloi2011,Martin2012,Bouche2012,DiamondStanic2012,
Leitherer2013,Chisholm2014}; see \citealt{Veilleux2005} for a review).
In principle, observations of outflows may be used to distinguish among outflow models, whether they are prescriptive (i.e.
sub-resolution) or predictive. However, inferring the total mass outflow rate is challenging, especially because the outflows
are multiphase \citep[e.g.][]{Martin2005,Martin2006,Leroy2015}.
Moreover, unless direct predictions for line profiles are provided by theorists, it is difficult
to ensure that the observational and theoretical definitions of what gas is outflowing are consistent.
Thus, it may not be possible to use current observations to distinguish amongst outflow models or provide phenomenological
scalings for simulations, but this is certainly a topic that is worthy of further consideration.

Because of resolution limitations, cosmological simulations typically incorporate
outflows driven by supernova feedback using sub-resolution models.
Some \citep[e.g.][]{NavarroWhite1993,SH03,OppenheimerDave2006,OppenheimerDave2008,Oppenheimer2010,
DallaVecchiaSchaye2008,Schaye2010,Hirschmann2013,Vogelsberger2013,AngelesAlcazar2014,Torrey2014}
treat supernova feedback by injecting kinetic energy (i.e. `kicks') into the neighboring gas elements. Many
of these implementations kinematically decouple the outflow particles to ensure that they escape
the ISM. The other commonly used technique is the `blastwave' model, in which thermal
energy is injected into the ISM surrounding supernova particles \citep[e.g.][]{Katz1996,ThackerCouchman2000,Stinson2006,
Stinson2013,Governato2007,Governato2010,Governato2012,DuboisTeyssier2008,Bournaud2010}.
However, because young star particles are typically located in regions of dense gas, where the cooling time is short,
such feedback is ineffective because the supernova feedback energy is rapidly radiated away.
To overcome this problem, it is common to either artificially disable cooling
for some fixed time interval \citep[e.g.][]{Stinson2006,Stinson2013} or store the supernova feedback energy until it is guaranteed to
generate a blastwave, albeit on artificially large (but numerically resolved) scales \citep[e.g.][]{DallaVecchiaSchaye2012,Crain2015,Schaye2015}.

Such treatments of outflows are useful because they enable these simulations to satisfy
a wide variety of observational constraints \citep[e.g.][]{OppenheimerDave2006,OppenheimerDave2008,
Oppenheimer2010,Oppenheimer2012,Dave2011b,Dave2011a,Dave2013,PuchweinSpringel2013,Vogelsberger2014,
Schaye2015,Ford2015},
although some significant discrepancies between the results of state-of-the-art cosmological simulations and models remain
(see \citealt{Vogelsberger2014}, \citealt{Genel2014}, and \citealt{Schaye2015} for discussions).
Similarly, SAMs must include strong outflows
to match observational constraints such as the galaxy stellar mass function \citep[e.g.][]{SomervillePrimack1999,Somerville2008,
Benson2003,Benson2014,Croton2006,Guo2011,Guo2013,Lu2014}.
Comparisons with observations of the circumgalactic and intergalactic media
\citep[e.g.][]{OppenheimerDave2009,Oppenheimer2012,Dave2010,Ford2013,Ford2014,Ford2015}
and X-ray observations \citep[e.g.][]{Cox2006,LeBrun2014,Zhang2014,Bogdan2015,Schaye2015}
can be used to constrain outflow models. However, because of their manner of implementation and
required tuning, these outflow models cannot be considered predictive, and it is highly desirable to
develop a \emph{physical, predictive} model for how stellar feedback drives outflows.

In SAMs \citep[e.g.][]{Croton2006,Somerville2008,Guo2011,Guo2013,White2015} and cosmological simulations
(\citealt{Okamoto2010,Vogelsberger2013,Torrey2014}; cf. e.g. \citealt{OppenheimerDave2008,Schaye2015}),
it is common to assume that the outflow mass-loading factor $\eta$ scales with the halo circular velocity
(see \citealt{Somerville2014} for additional discussion).
Thus, at fixed halo mass (and thus approximately constant stellar mass because the redshift evolution in the stellar mass--halo mass relation is weak;
\citealt{Behroozi2013}), the mass-loading factor decreases with increasing redshift. For Milky Way-mass
galaxies, this effect inevitably leads to too much high-redshift star formation or over-suppression of low-redshift
star formation, depending on the assumed normalisation of the $\eta-\vchalo$ relation \citep[e.g.][]{Torrey2014,White2015}.
Moreover, observational constraints require that the mass-loading factor
at fixed stellar mass \emph{increases} with increasing redshift \citep{Mitra2015}; this trend cannot be reproduced
in models that assume simple $\eta \propto \vchalo^{-1}$ or $\eta \propto \vchalo^{-2}$ scalings. An alternative
method for calculating the mass-loading factor is thus required.

Recently, multiple groups have made significant advances by generating outflows self-consistently
via stellar feedback without resorting to delayed cooling or kinematic decoupling
\citep{Hopkins2012winds,Hopkins2013merger_winds,Hopkins2014,Agertz2013,Agertz2015a,Agertz2015b}.
In particular, \citet{Muratov2015} have demonstrated that stellar feedback can effectively drive
outflows in galaxies of stellar mass $\mstar \la 10^{10} \msun$
at all redshifts. The simulated galaxies agree well with observational constraints, such as 
the stellar mass--halo mass, Kennicutt--Schmidt,
and SFR--stellar mass relations \citep{Hopkins2014}.
The \emph{predicted} mass-loading factors are considerably less
than typically assumed in large-volume cosmological simulations and SAMs, which suggests
that the latter use outflow models that are inconsistent with the scalings of
\citet{Muratov2015} or/and they must invoke higher mass-loading factors to compensate for limitations
of the (sub-resolution) feedback models employed.
Moreover, \citet{Muratov2015} noted that outflows were significantly suppressed
in their simulated galaxies with $\mstar \ga 10^{10} \msun$ at low redshift
($z \la 1$). This suppression of outflows enables these massive galaxies to transition from
highly turbulent, clumpy discs characterised by bursts of star formation and subsequent outflows
to well-ordered, thin, steadily star-forming disc galaxies at $z = 0$.

The aforementioned simulations have the advantage of including stellar feedback in an explicit,
resolved manner and generating outflows self-consistently. However, it can be difficult to extract
physical explanations for phenomena `observed' in such simulations because of the complexity
of the simulations. For this reason, \citet{Muratov2015} did not identify the physical origin of the
suppression of outflows in their simulated $\mstar \ga 10^{10} \msun$ galaxies at $z \la 1$.
Simple analytic models are a useful complementary tool that can be used to
posit potential physical mechanisms, the veracity of which can be checked via detailed simulations
and, ideally, comparisons with observations. In this work, we present an analytic theory for
stellar-feedback-driven galactic outflows. One of the most interesting features of the model is that, as
we will discuss in detail below, it
predicts suppression of outflows in massive galaxies at low redshift that is qualitatively
consistent with the simulation-based results of \citet{Muratov2015}. Consequently, our model may
offer a physical explanation for those results.

This work builds upon a number of previous works. The first key assumption of our model is that
star formation is self-regulated. By this, we mean that momentum deposition from
stellar feedback or photo-heating from star formation
is assumed to provide vertical pressure support against gravity such that a galaxy
maintains a quasi-equilibrium state.
Under this assumption, the rate of momentum or energy deposition sets the star formation rate.
Various previous works have presented such models \citep[e.g.][]{Murray2005,Murray2011rad,Thompson2005,
Ostriker2010,OstrikerShetty2011,ShettyOstriker2012,FQH2013}. In particular, we rely heavily on the work of
\citet[][hereafter FQH13]{FQH2013}.

The second key assumption is that the gas that will be blown out by stellar feedback corresponds to
the gas that can be accelerated to the local escape velocity before the density distribution is reset
by turbulence. As we demonstrate below, this implies that gas below some maximum surface density
will be blown out. Because the ISM is turbulent, a given `macroscopic' patch of ISM exhibits sub-patches
with a range of surface densities. \citet[][hereafter TK14]{TK2014} demonstrated that this property of
supersonic turbulence implies that even when a galaxy is globally below the Eddington limit (i.e. the
surface density above which radiation pressure on dust provides the dominant pressure
support against gravity; \citealt{Scoville2003,Thompson2005}),
it will have sub-patches that are super-Eddington. As noted by TK14, virtually all previous models for
how momentum deposition drives turbulence (e.g. \citealt{KrumholzMcKee2005,Krumholz2009,
Murray2010,OstrikerShetty2011}; FQH13) and generates outflows
\citep[e.g.][]{Murray2005,Murray2011rad,Thompson2005}
have considered only the mean gas surface density. We build upon the work of TK14 by applying an
analysis similar to theirs to a self-regulated galaxy formation model. Specifically, we use the
results of FQH13 for the high-surface-density regime, in which turbulent pressure dominates.
For the low-surface-density regime, in which thermal pressure dominates, we calculate the
quasi-equilibrium SFR surface density relation (see \citealt{Ostriker2010} for a more detailed
treatment). Through this synthesis, we predict the mass-loading factor of stellar-feedback-driven outflows
as a function of galaxy properties.

The remainder of this work is organised as follows: \sref{S:f_w} presents our method for calculating
the mass outflow rate for momentum-driven outflows in a turbulent ISM. \sref{S:turb} derives
expressions for the mass-loading factor in the regime in which a galaxy/patch is supported
by turbulent pressure, whereas \sref{S:therm} considers the regime in which thermal pressure
balances gravity. In \sref{S:eta_global_props}, we use empirically based scalings 
to demonstrate how the mass-loading factor depends on stellar mass and redshift.
\sref{S:implications} presents some implications
of our model for galaxy evolution, and \sref{S:implementation} outlines how one could implement
our model in hydrodynamical simulations or SAMs. We summarize our conclusions in \sref{S:conclusions}.
In \aref{S:therm_app}, we derive the equilibrium SFR surface density and mass-loading factor relations for the thermal
pressure-dominated regime. \aref{S:galaxy_scaling_relations} presents the empirically based scaling relations
that we use to determine the dependence of the mass-loading factor on stellar mass and
redshift in \sref{S:eta_global_props}.
In \aref{S:regimes}, we use these relations to determine which regime is relevant as a function of mass and redshift.

\section{How stellar feedback drives outflows in a turbulent ISM} \label{S:f_w}

In this section, we present our approach for calculating the mass outflow rate of an outflow driven
by momentum deposition from stellar feedback. Crucially, we consider a turbulent ISM
in which a patch/disc\footnote{We will use the terms `patch' and `disc' interchangeably because the model
can be applied to either an entire galaxy, using galaxy-averaged quantities, or sub-regions of galaxies,
as long as the latter are of a size greater than the disc scaleheight.} of some mean gas surface density contains sub-patches with a range of
gas surface densities.\footnote{Because we are considering both the mean gas surface density of a
patch and the surface densities of sub-patches, we distinguish the former using the notation
$\msiggas$. All other quantities, such as the equilibrium SFR surface density, correspond to
averages over the total patch/disc. For simplicity of notation, we omit the angle brackets for all
spatially averaged quantities except $\msiggas$.}
We first present our criterion for determining which gas will be blown out on a coherence time
(i.e. before turbulence `resets' the density distribution), and we then
discuss how we calculate the fraction of the ISM mass that satisfies this criterion.

\subsection{Which gas can be blown out?}

As noted by TK14, in a turbulent ISM, a given patch of ISM with mean gas surface density
$\langle \siggas \rangle$ will contain sub-patches with a range of surface densities.
For a sub-patch of ISM with gas surface density $\siggas$
to be expelled, \rev{we assume that} it must be accelerated to the escape velocity of the disc, $\vesc$, within
a coherence time. \rev{The motivation for this assumption is that over a coherence time, turbulence
will `reset' the gas surface density distribution of the disc; for patches with surface density initially
less than the mean surface density, this will tend to cause the surface density below the patch to increase,
and this high-surface-density gas will `block' further momentum injection from local stellar feedback.
Consequently, the patch will no longer be accelerated and will rain back down onto the disc rather than be
expelled as an outflow. However, this assumption may not hold because driving outflows may be a relatively
slow process that is the result of several acceleration steps \citep[e.g.][]{Hill2012,Girichidis2016}.
Local stellar feedback from e.g. supernovae may push a patch to some height
above the disc but not accelerate it to the escape velocity. Then, other sources of acceleration, such
as supernovae from runaway OB stars \citep[e.g.][]{Li2015} or cosmic rays \citep{Uhlig2012,Booth2013,Hanasz2013,
Salem2014a,Salem2014b,Girichidis2016}, may provide the remaining momentum deposition necessary to accelerate
the patch to the escape velocity. In principle, we could crudely account for such multi-step acceleration
in our model by requiring the gas to be accelerated only to some fraction of the escape velocity.
However, given the poorly understood nature of the above processes and considerations of simplicity,
we will not do so in the present work.}

\rev{With the above caveats in mind, to make the problem tractable, we will assume that a sub-patch of ISM must be accelerate to the
escape velocity within a coherence time, and we recognise that this assumption may require revision
in future work.}
In turbulence, the coherence time is $\sim t_{\rm eddy} \sim \sigma_{\rm eddy}/l_{\rm eddy}$, where $t_{\rm eddy}$
and $\sigma_{\rm eddy}$ are the characteristic timescale and turbulent velocity associated with eddies on a spatial scale
$l_{\rm eddy}$. In a vertically
stratified, supersonically turbulent disc with scaleheight $h$, vertical force balance requires that
the largest eddies have scaleheight $\sim h \sim \sigt/\Omega$,
where $\Omega$ is the local orbital frequency and $\sigt$ is the characteristic
turbulent velocity of the medium for eddies with $l_{\rm eddy} \sim h$.
Since $\siggas$ is the gas volume density integrated in the vertical direction, we are considering \emph{only} patches
integrated through the disc
(i.e. with minimum individual size $\sim h$). Moreover, because the large scales contain all the power in density
fluctuations, we are primarily concerned with eddies with $l_{\rm eddy} \sim h$. Thus, we obtain the typical coherence
time $\sim \Omega^{-1}$, and we therefore require
\begin{equation}
\dot{P} \Omega^{-1} > m \vesc,
\end{equation}
where $\dot{P}$ is the momentum deposition rate and $m$ is the mass of the patch of ISM.
For a disc in radial centrifugal balance in an isothermal potential with circular velocity $\vc$,
the total mass inclosed with radius $r$ is $M(r) = \vc^2 r/G$.
The escape velocity is $\vesc = \sqrt{G M(r)/r} = \sqrt{2} \vc$. Thus,
\begin{equation}
\dot{P} \ga \sqrt{2} \Omega m \vc.
\end{equation}
In terms of surface density,
\begin{equation} \label{eq:p_requirement}
\sigpdot \ga \sqrt{2} \Omega \siggas \vc,
\end{equation}
where $\sigpdot$ is the momentum deposition rate per unit area.

For sources of stellar feedback such as winds, supernovae, and radiation pressure,
the momentum deposition rate per unit area is given by
\begin{equation} \label{eq:p-sfr}
\sigpdot = \left(\frac{P_{\star}}{m_{\star}}\right) \sigsfr,
\end{equation}
where $\sigsfr$ is the mean SFR surface density of the patch
and $(P_{\star}/m_{\star})$ is the momentum deposited per unit stellar mass formed. This value varies
depending on the source of stellar feedback being considered. Following FQH13, we will use
$(P_{\star}/m_{\star}) = 3000$ km s$^{-1}$, which is appropriate for supernova feedback (see FQH13 and references therein),
as our fiducial value. We will retain this factor such that our results can be generalized for other sources of stellar feedback.

Thus, the surface density below which gas will be accelerated to the escape velocity on a coherence time
and thus blown out is
\begin{equation} \label{eq:siggas_max_1}
\siggasmax \equiv \left(\frac{P_{\star}}{m_{\star}}\right) \frac{\sigsfr}{\sqrt{2} \Omega \vc}.
\end{equation}

The above framework assumes that the outflows are momentum-driven. The
fiducial value $P_{\star}/m_{\star} = 3000 \kmpers$ is based on the assumption that SN remnants
go through the energy-conserving Sedov-Taylor \citep{Taylor1950,Sedov1959} phase in which energy is converted
into momentum. The reason that the outflows can be considered momentum-driven is that we assume
that by the time the outflows `break out' of the disc, the energy-conserving phase has ended (see FQH13 for details).
This assumption is justified as long as the cooling radii of the supernova ejecta are less than the disc scaleheight.
If this condition does not hold (i.e. a supernova remnant expands to of order the disc scaleheight or greater
while still in the Sedov-Taylor phase), the outflows should be treated as energy-conserving.
We can estimate when this assumption will be violated as follows. The cooling radius for an individual supernova remnant is
$R_{\rm cool} \sim 14 ~\mathrm{pc} (n/\mathrm{cm}^{-3})^{-3/7}$ \citep{Cioffi1988,Thornton1998},
where $n$ is the number density of the ambient medium;
we have ignored the weak metallicity dependence. The disc scaleheight is $h \sim \cs/\Omega$, where
$\cs$ is the sound speed.\footnote{This expression for the disc scaleheight assumes that the disc is supported by thermal rather than
turbulent pressure. Below, we shall see that this is the case for the relevant surface densities.}
Using the relation $n \sim \msiggas/2m_{\rm p}h$, where $m_{\rm p}$ is the proton mass,
the cooling radius can be expressed as 
\begin{equation}
R_{\rm cool} \sim 14 ~\mathrm{pc} \left(\frac{\msiggas/2m_{\rm p}h}{1 ~\mathrm{cm}^{-3}}\right)^{-3/7}.
\end{equation}
The cooling radius is greater than the disc scaleheight, $R_{\rm cool} \ga h$, when
\begin{equation}
\msiggas \la 1 \msun ~\mathrm{pc}^{-2} \left(\frac{h}{14 ~\mathrm{pc}}\right)^{-4/3}.
\end{equation}
Thus, we expect energetically driven outflows to be significant only when the local gas surface density
(i.e. the gas surface density averaged over the cooling radius) is $\la 1 \msun$ pc$^{-2}$. Because
patches with such low surface density will almost always be below our calculated critical surface density,
we will include this material in the outflowing material, and the actual driving method is irrelevant
for the purposes of calculating the outflow fraction.

It is certainly possible for this condition to be satisfied in the outer regions of discs or if the surface
density of the local ISM has already been reduced considerably by other forms of feedback or other
supernovae before a supernova explodes \citep[e.g.][]{Larson1974,Canto2000,WadaNorman2001,
Martizzi2015}. Still, even if it is possible
for supernova ejecta to escape the galaxy while still in the energy-conserving phase, the expected
mass loading is factor is small because little material will be entrained; thus, the amount of outflowing
gas will be of order the supernova ejecta mass.
For standard initial mass functions, on average, each supernova ejects of order 10 $\msun$, and there is approximately one
supernova per $100 \msun$ formed \citep{Kroupa2001,Chabrier2003}. Thus, we expect the mass-loading factor to be $\la 0.1$, and indeed,
numerical simulations of energy-driven outflows from supernovae suggest that very little mass is
blown out \citep[e.g.][]{MacLowFerrara1999,StricklandStevens2000}.
For this reason, we neglect this form of feedback in the current work.

\subsection{What fraction of the ISM will be blown out?}

We now calculate the fraction of the ISM with $\siggas < \siggasmax$ by considering the distribution of gas surface density
in a turbulent ISM, closely following the analysis of TK14.
For supersonic isothermal turbulence, the probability distribution functions (PDFs) of both the volume and surface densities are approximately
lognormal \citep[e.g.][]{Ostriker2001,VS2001,Federrath2010,Konstandin2012a,Konstandin2012b,Konstandin2015}.
Given a value of $\siggasmax$, we can use the gas density PDF to
calculate the mass fraction of the gas that is blown out by stellar-feedback-driven outflows,
$\fw$, which we refer to as the `outflow fraction', by computing the mass fraction of the disc that has $\siggas < \siggasmax$.
This can be done by combining equations (8), (12) and (14) of TK14. Define
\begin{equation}
\xw \equiv \ln\left(\frac{\siggasmax}{\langle \siggas \rangle}\right).
\end{equation}
Then, from equation (8) of TK14, the outflow fraction is
\begin{equation} \label{eq:f_w}
\fw = \int_{-\infty}^{\xw} p_-(x) dx = \frac{1}{2} \left[1 - \mathrm{erf} \left( \frac{-2 \xw + \sigma^2_{\ln \siggas}}{2 \sqrt{2}
\sigma_{\ln \siggas}}\right)\right],
\end{equation}
where erf denotes the error function and $\sigma_{\ln \siggas}$ is the dispersion of the logarithm of the gas surface
density. Note that we have used $p_-(x)$ because we wish to calculate the fraction of the mass (rather than the area) that has
$\siggas < \siggasmax$; see TK14 for details.
The width of the PDF of the logarithm of the gas surface density is (equation 12 of TK14)
\begin{equation} \label{eq:tk14_eq12}
\sigma^2_{\ln \siggas} \approx (1+ R \mathcal{M}^2/4),
\end{equation}
where $\mathcal{M} \equiv \sigt/\cs$ is the Mach number of the turbulence and R is given by equation (14) of TK14,
\begin{equation} \label{eq:tk14_eq14}
R = \frac{1}{2} \left( \frac{3-\alpha}{2-\alpha} \right) \left[ \frac{1-\mathcal{M}^{2(2-\alpha)}}{1-\mathcal{M}^{2(3-\alpha)}} \right].
\end{equation}
Following TK14, we assume that the power-law index of the power spectrum of the turbulence is $\alpha = 2.5$
because in the turbulent-pressure-supported regime, the turbulence is generally highly supersonic.
When our model predicts that the turbulence is trans-sonic, we instead use $\alpha = 3.7$.
However, our results are insensitive to the exact value of $\alpha$ used.
The above equations can be used to calculate $\fw$ if $\mathcal{M}$ and $\siggasmax$ are known.

The primary quantity with which we are concerned in this work is the mass-loading factor, $\eta \equiv \mwind/\sfr = \sigwind/\sigsfr$,
where $\mwind$ is the mass outflow rate and $\sigwind$ is the mass outflow rate per unit area, which is $\sigwind = \fw \langle \siggas \rangle \Omega$
by construction in our model. Thus, the mass-loading factor is
\begin{equation} \label{eq:eta}
\eta = \frac{\fw \langle \siggas \rangle \Omega}{\sigsfr}.
\end{equation}

\section{Turbulent pressure-supported regime} \label{S:turb}

Now that we have presented the formalism for calculating the mass outflow rate, we must determine the
equilibrium SFR surface density relations.
We first consider the limit in which turbulent pressure is the dominant source
of pressure support against gravity. This regime is studied in detail in FQH13 (see also \citealt{Thompson2005}
and and \citealt{OstrikerShetty2011}), so we will only summarise the model here.
Stellar feedback injects momentum into the ISM. A fraction $\fw$ drives outflows, whereas the remaining
$(1-\fw)$ pushes on gas that will not be blown out and drives turbulence with a momentum injection rate
per unit area of $(1-\fw) \sigsfr (P_{\star}/m_{\star})$. We have included the $(1-\fw)$ factor so that we do not
double-count the momentum that drives outflows rather than stirring the gas. As argued above,
turbulence dissipates energy on a crossing time, $\Omega^{-1}$. In equilibrium, the momentum
injection rate per unit area must equal the dissipation rate per unit area, $\sim (1-\fw) \msiggas \sigt \Omega$.
We also include the $(1-\fw)$ factor here because we must balance the dissipation only in the gas that 
remains in the disc, not the outflowing material. Setting the momentum injection and dissipation rates
equal, we have
\begin{equation} \label{eq:sig_sfr_from_mom_equality}
\sigsfr \approx \left(\frac{P_{\star}}{m_{\star}}\right)^{-1} \sigt \Omega \msiggas.
\end{equation}
The \citet{Toomre1964} $Q$ parameter is
\begin{equation} \label{eq:q}
Q = \frac{\kappa \sqrt{\cs^2 + \sigt^2}}{\pi G \msiggas},
\end{equation}
where $\kappa \sim \sqrt{2} \Omega$ is the epicyclic frequency, $\cs$ is the sound speed, $\sigt$ is
the turbulent velocity dispersion and
$\msiggas$ is the mean gas surface density of the patch/disc.
We assume that turbulence provides the dominant vertical pressure support against gravity;
thus, $\sqrt{\cs^2 + \sigt^2} \sim \sigt$.
We can use the Toomre Q to rewrite equation (\ref{eq:sig_sfr_from_mom_equality})
in the following form:
\begin{equation}
\sigsfr \approx \frac{\pi G Q}{\sqrt{2}} \left(\frac{P_{\star}}{m_{\star}}\right)^{-1} \msiggas^2.
\end{equation}

FQH13 present a somewhat more sophisticated derivation of the equilibrium SFR surface density relation
in this regime. Their expression has the same dependences on $\msiggas$, $Q$ and $(P_{\star}/m_{\star})$,
but the prefactor differs. Below, we will use the expression derived by FQH13 (their equation 18):
\begin{equation} \label{eq:cafg_eq18}
\sigsfrturb = \frac{2 \sqrt{2} \pi G Q \phi}{\mathcal{F}'} \left(\frac{P_{\star}}{m_{\star}}\right)^{-1}
\langle \siggas \rangle^2,
\end{equation}
where $G$ is the gravitational constant, $Q$ is the Toomre $Q$ parameter, $\phi$ is an order-unity parameter
that is $\sim 1$ for a thin disc in a spherical potential and $\sim 1/Q$ for a self-gravitating pure gas disc,
and $\mathcal{F}'$ is a dimensionless parameter that encapsulates uncertain factors that are of order
unity (see \citealt{Thompson2005} and \citealt{OstrikerShetty2011} for similar expressions).\footnote{In their equation (18), FQH13
include the term $(1-f_{\mathrm{w}})$, where $f_{\mathrm{w}}$ is the fraction of the input momentum
that drives outflows (rather than supports the disc against gravity); this is equivalent to our $\fw$.
However, as we note above, the outflowing gas does not need to be
supported against gravity; thus, there should be a $(1-f_{\mathrm{w}})$ factor on the right-hand side of their equation (14), which
would eliminate the $(1-f_{\mathrm{w}})$ term in their equation (18). Consequently, we use the prefactor
$\mathcal{F}' = f_{\rm P} f_{\rm h} \gamma$, where $f_{\rm P}$, $f_{\rm h}$, and $\gamma$ are order-unity
coefficients that encapsulate various uncertainties (see FQH13 for details), and we note that this factor does not include the
$(1-f_{\mathrm{w}})$ term.}
We have compared this equilibrium SFR surface density relation with the results of high-resolution galaxy simulations
that include resolved stellar feedback and found excellent agreement
(see also \citealt{KOK2013}).

Given the equilibrium SFR surface density, we can predict how the outflow fraction
depends on galaxy properties. Combining equations (\ref{eq:siggas_max_1}) and (\ref{eq:cafg_eq18}),
we find that in this regime, the critical surface density below which gas can be blown out on a coherence time is
\begin{equation} \label{eq:sig_max}
\siggasmax = \frac{2 \pi G Q \phi}{\mathcal{F}' \Omega \vc} \langle \siggas \rangle^2 \equiv \siggasmaxturb.
\end{equation}

We can express $\siggasmaxturb$ in a more useful form as follows. 
Using the definition of the Toomre Q (equation \ref{eq:q}) and $\sigma_{\mathrm{T}}^2 \gg \cs^2$
(because we are considering the limit in which turbulence dominates the pressure support),
we have
\begin{equation} \label{eq:omega_over_sig_gas}
\frac{\langle \siggas \rangle}{\Omega} \sim \frac{\sqrt{2} \sigma_{\mathrm{T}}}{\pi G \qturb},
\end{equation}
where $\qturb$ denotes the turbulent Toomre $Q$ (i.e. equation \ref{eq:q} with $\cs = 0$).
Using equation (\ref{eq:omega_over_sig_gas}), equation (\ref{eq:sig_max}) can be rewritten as
\begin{equation} \label{eq:sig_gas_max_2}
\siggasmaxturb = \frac{2 \sqrt{2} \pi \phi}{\mathcal{F}'} \frac{\sigt}{\vc} \langle \siggas \rangle.
\end{equation}
For a turbulent-pressure-supported disc (equation 9 of FQH13),
\begin{equation} \label{eq:sig_t_2}
\frac{\sigma_{\mathrm{T}}}{\vc} = \frac{\qturb}{\sqrt{2}} \fg,
\end{equation}
where we have used $\vc = \sqrt{2} \sigma$, in which $\sigma$ is the velocity dispersion of the isothermal potential.
In the above equation, $\fg \equiv \mgas/(\mgas + \mstar)$ is the gas fraction of the disc/patch; note that $\fg \le 1$
by definition. Combining equations (\ref{eq:sig_gas_max_2}) and (\ref{eq:sig_t_2}), we have
\begin{eqnarray}
\siggasmaxturb &=& \frac{2 \phi \qturb}{\mathcal{F}'} \fg \langle \siggas \rangle \label{eq:sigmax_turb_final} \\
&=& \frac{2 \sqrt{2} \phi}{\mathcal{F}'} \left(\frac{\sigt}{\vc}\right) \langle \siggas \rangle.
\end{eqnarray}
Thus,
\begin{eqnarray}
\xwturb &=& \ln \left(\frac{2 \phi \qturb}{\mathcal{F}'} \fg\right) \label{eq:xw_turb} \\
&=& \ln \left[ \frac{2 \sqrt{2} \phi}{\mathcal{F}'} \left(\frac{\sigt}{\vc}\right) \right].
\end{eqnarray}

To calculate the Mach number, we require the sound speed,
\begin{equation} \label{eq:cs}
\cs = \sqrt{\frac{k_B T}{\mu}},
\end{equation}
where $\mu \approx 0.6$ amu for a fully ionized plasma of primordial composition.
Thus, for $T = 10^4$ K, $\cs \approx 12 \kmpers$.
Equations (\ref{eq:sig_t_2}) and (\ref{eq:cs}) imply
\begin{eqnarray} \label{eq:mach_num_turb}
\mathcal{M} &=& \frac{\qturb \fg \vc}{\sqrt{2 k_{\mathrm{B}} T/\mu}} \equiv \mturb \\
&\approx& 6 \qturb \left(\frac{\mu}{0.6}\right)^{1/2} \left(\frac{T}{10^4 ~\mathrm{K}}\right)^{-1/2} \left(\frac{\fg \vc}{100 \kmpers}\right)
\label{eq:M}.
\end{eqnarray}
We assume that the diffuse, turbulence-supported ISM has a temperature of
$\sim 10^4$ K.\footnote{For $T = 10^4$ K, equation
(\ref{eq:mach_num_turb}) yields $\mathcal{M} < 1$ for $\fg \vc < 17 \kmpers$. However, gas cooling onto galaxies of this
circular velocity should be suppressed by photoionisation from the metagalactic ultraviolet background \citep{ThoulWeinberg1996,
Bullock2000,Hoeft2006,Okamoto2008}. Thus, galaxies with $\vc \la 20 \kmpers$ (equivalent to $\mstar \la 10^6 \msun$)
are not considered in detail in this work.}

\begin{figure}
\centering
\includegraphics[width=0.9\columnwidth]{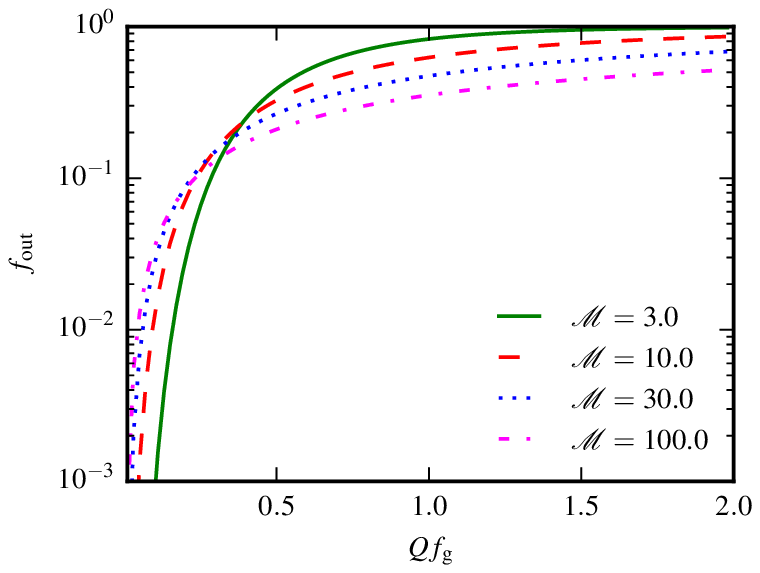}
\caption{The fraction of the ISM mass that will be blown out on a coherence time, $\fwturb$,
versus the product of the turbulent Toomre Q and the gas fraction, $\qturb \fg$ (= $\sqrt{2} \sigt/\vc$),
for different Mach numbers (see the legend); we assume that the order-unity parameters $\phi$ and $\mathcal{F}'$ are exactly unity.
There is a critical $\qturb \fg$ value below which $\fwturb$ decreases exponentially because for
low $\qturb \fg$, the rms turbulent velocity is much less than the escape velocity. Thus, only extremely low-surface-density
patches, which lie outside of the `core' of the lognormal distribution, can be blown out (because the momentum
deposition rate depends on the \emph{mean} SFR and thus gas surface density, whereas the amount of momentum required
to accelerate a patch to the escape velocity depends on the \emph{local} gas surface density).
When $\qturb \fg$ is large, the rms turbulent velocity is sufficiently high that patches with surface densities
of order the mean surface density can be blown out, and the outflow fraction becomes of order unity.}
\label{fig:fw_vs_qfg}
\end{figure}

\fref{fig:fw_vs_qfg} shows the outflow fraction ($\fw$, i.e. the solution to equation \ref{eq:f_w})
versus $\qturb \fg$ for various Mach numbers (see the legend).
$\fw$ increases with $\qturb \fg = \sqrt{2} \sigt/\vc$ because the ratio $\siggasmax/\langle \siggas \rangle \propto \qturb \fg$. 
Moreover, there is a sharp cutoff at $\qturb \fg \sim 0.3$; this cutoff occurs at lower $\qturb \fg$ values for higher Mach numbers
because the lognormal gas surface density PDF broadens as the Mach number is increased (see below for details).
The reason for this behavior is as follows: $\siggasmaxturb$, the surface density below which gas is blown out, scales as
$\sigsfr$ (equation \ref{eq:siggas_max_1}), which scales as $\qturb \msiggas^2$ (equation \ref{eq:cafg_eq18}).
Thus, the ratio $\siggasmaxturb/\msiggas \propto \qturb \msiggas$.
As $\qturb \fg$ decreases, $\qturb \msiggas$
decreases, and thus $\siggasmaxturb/\msiggas$ decreases. Consequently, increasingly less of the density PDF
is sampled as $\fg$ decreases, and thus $\fw$ decreases (see equation \ref{eq:f_w}).\footnote{Note that $\fw$ will
increase with increasing $\msiggas$ as long as $\sigsfr$ depends super-linearly on $\msiggas$.}
The cutoff occurs when
$\siggasmaxturb/\langle \siggas \rangle$ is sufficiently small that only the low-surface-density tail of the lognormal PDF
is sampled.

Put more physically, when $\qturb \fg$ is low, the rms turbulent velocity is much less than the escape velocity.
The amount of momentum deposited into a patch on a coherence time, and thus the rms turbulent velocity,
depends on the \emph{mean} SFR and thus gas surface density.
In contrast, the amount of momentum required to accelerate a patch to the escape velocity
on a coherence time depends on the \emph{local} gas surface density (i.e. for a fixed amount of momentum,
the typical velocity to which gas is accelerated decreases as the mass of gas being pushed is increased).
Thus, when the rms turbulent velocity is much less than the escape velocity, only extremely low-surface-density
patches (relative to the mean surface density) can be accelerated to the escape velocity. As $\qturb \fg$ is increased,
the rms turbulent velocity becomes comparable to the escape velocity. In this case, patches with surface densities of order the mean surface
density can be accelerated to the escape velocity, and the outflow fraction becomes of order unity.

The dependence of the outflow fraction on the Mach number of the turbulence, $\mturb$, evident in
\fref{fig:fw_vs_qfg} is also of interest. As the Mach number is increased, the lognormal PDF broadens
(see fig. 1 of TK14): the tails contain
an increased fraction of the ISM mass, and the core of the distribution contains less (i.e. the fraction of the ISM
with surface density near the mean decreases). Consequently, when the critical surface density is in the low-surface-density
tail of the lognormal distribution ($\qturb \fg \la 0.5$ because when $\qturb \fg = 0.5$, $\siggasmax = \msiggas$),
increasing the Mach number increases the outflow fraction.
Conversely, when the critical surface density is of order the mean surface density (i.e. the rms turbulent velocity
is of order the escape velocity), the outflow fraction decreases with increasing Mach number because an
increased fraction of the mass has surface density significantly greater than the mean and thus cannot be blown out.

We have determined the critical surface density below which gas can be blown out on a coherence time.
We are thus now in a position to calculate the mass-loading factor, $\eta$ (equation \ref{eq:eta}). For clarity, we use the notation $\etaturb$
to remind the reader that the below expressions apply in the turbulent-pressure-supported regime.
Using the equilibrium SFR surface density relation for this regime (equation \ref{eq:cafg_eq18}),
we have
\begin{eqnarray}
\etaturb &=& \frac{\mathcal{F}'}{2 \sqrt{2} \pi G \qturb \phi} \fw \left(\frac{P_{\star}}{m_{\star}}\right) \Omega \langle \siggas \rangle^{-1}
\label{eq:eta_omega_siggas} \\
& \approx& 80 \frac{\mathcal{F}'}{\qturb \phi} \fw \left(\frac{P_{\star}/m_{\star}}{3000 \kmpers}\right) \\
&\times& \left(\frac{\Omega}{10 ~\mathrm{Gyr}^{-1}}\right) \left(\frac{\langle \siggas \rangle}{10^3 \msun ~\mathrm{pc}^{-2}}\right)^{-1}.
\nonumber
\end{eqnarray}
Equation (\ref{eq:eta_omega_siggas}) can be recast into another useful form using equation (\ref{eq:omega_over_sig_gas}):
\begin{eqnarray}
\etaturb &=& \frac{\mathcal{F}'}{2 \sqrt{2} \phi} \fw \left(\frac{P_{\star}}{m_{\star}}\right) \sigt^{-1} \label{eq:eta_sigt} \\
&\approx& 10^2 \frac{\mathcal{F}'}{\phi} \fw \left(\frac{P_{\star}/m_{\star}}{3000 \kmpers}\right) \left(\frac{\sigt}{10 \kmpers}\right)^{-1}.
\end{eqnarray}
A third useful expression for $\eta$ can be obtained using equations (\ref{eq:sig_t_2}) and (\ref{eq:eta_sigt}):
\begin{eqnarray}
\etaturb &=& \frac{\mathcal{F}'}{2 \qturb \phi} \fw \left(\frac{P_{\star}}{m_{\star}}\right) (\fg \vc)^{-1} \\
&\approx& 15 \frac{\mathcal{F}'}{\qturb \phi} \fw \left(\frac{P_{\star}/m_{\star}}{3000 \kmpers}\right) \label{eq:eta_fg_vc}\\
&\times& \left(\frac{\fg \vc}{100 \kmpers}\right)^{-1}. \nonumber
\end{eqnarray}
It is important to keep in mind that in the above expressions for $\eta$, $\fw$ depends on the galaxy properties
($\mathcal{M}$ and the product $\qturb \fg$). Thus, for example, the scaling $\eta \propto (\fg \vc)^{-1}$ will not hold
if $\fw$ varies strongly with $\fg \vc$.

\begin{figure}
\centering
\includegraphics[width=0.9\columnwidth]{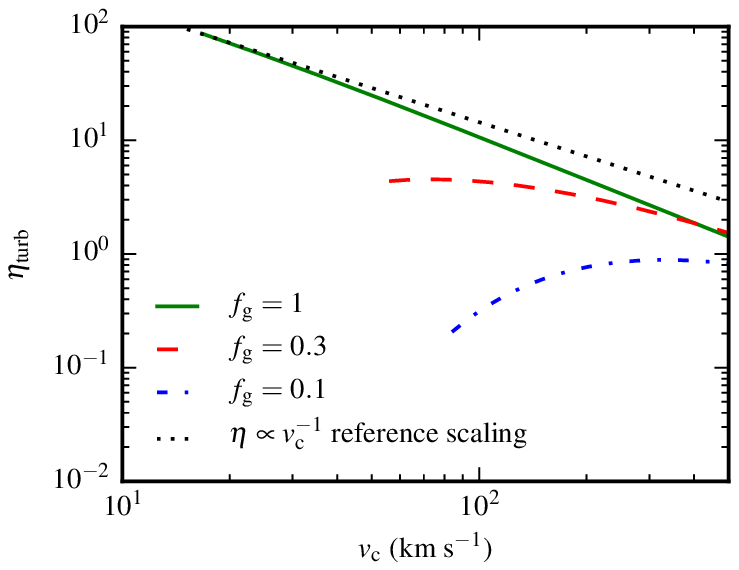}
\caption{The mass-loading factor ($\etaturb$), in the turbulent-pressure-supported regime versus the galaxy circular
velocity ($\vc$) for different values of the gas fraction (see the legend).
The reference scaling for a momentum-driven outflow, $\etaturb \propto \vc^{-1}$ (with arbitrary normalization), is
indicated by the black dotted line. When the gas fraction is high,
$\etaturb$ decreases slightly more steeply with $\vc$ than the $\vc^{-1}$ reference scaling because $\fwturb$ (not shown)
decreases mildly with $\vc$. When the gas fraction is near or below the critical value of $\sim 0.3$, $\etaturb$ is suppressed
considerably. The suppression is stronger at lower $\vc$ values because when $\siggasmax$, the critical surface density below
which gas will be blown out on a coherence time, is in the low-surface-density tail of the PDF,
the outflow fraction, $\fw$, depends very sensitively on the Mach number, $\mathcal{M}$ (see \fref{fig:fw_vs_qfg}).
For fixed $\fg$, lower $\vc$ results in lower $\mathcal{M}$
and thus significantly lower $\fw$. (N.B. The red dashed and blue dot-dashed lines are truncated at the $\vc$ values at which $\mathcal{M} = 1$
because galaxies with lower $\vc$ values would be supported by thermal, not turbulent, pressure.)}
\label{fig:eta_vs_vc}
\end{figure}

We stress that $\vc$ is the circular velocity of the galaxy at the effective radius,
not the circular velocity of the halo. At fixed galaxy mass,
$\vc$ evolves only weakly with redshift (i.e. the Tully-Fisher relation is approximately redshift-independent; e.g. \citealt{Miller2011,Miller2012,Miller2013}).
In contrast, both the circular velocity at the virial radius, $V_{\rm vir}$, and the maximum circular velocity of the halo, $V_{\rm max}$,
evolve with redshift: at fixed halo mass, both $V_{\rm vir}$ and $V_{\rm max}$ decrease with $z$ \citep[e.g.][]{SomervillePrimack1999,
Bullock2001}. Thus, if $\eta$ is calculated using $V_{\rm vir}$ or $V_{\rm max}$ rather than the circular velocity
of the galaxy, $\vc$, at fixed virial mass (and thus approximately constant galaxy mass because the stellar-to-halo mass ratio
evolves weakly with redshift; \citealt{Behroozi2013,Moster2013}), $\eta$ will decrease with redshift. In contrast, when the circular
velocity at the galaxy, $\vc$, is used, at fixed galaxy mass, $\eta$ is approximately independent of redshift (as long as we are not in the regime in
which $\fw$ is exponentially suppressed), as we shall see below. Thus, when implementing our model, it is crucial that the mass loading
factor be calculated using the galaxy circular velocity. If a halo property must be used, it is better to use the maximum circular velocity of the
halo because at fixed $M_{\rm vir}$, the dependence of $V_{\rm max}$ on redshift is significantly weaker than that of $V_{\rm vir}$
\citep{Bullock2001}.

\begin{figure}
\centering
\includegraphics[width=0.9\columnwidth]{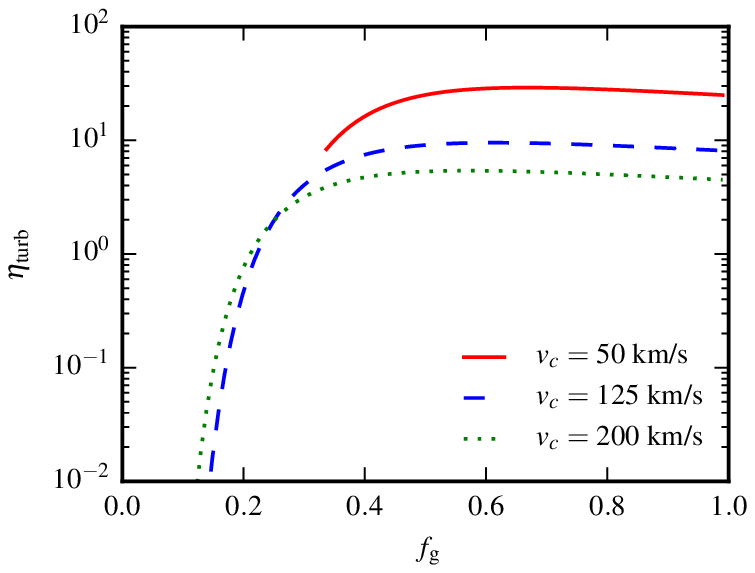}
\caption{The mass-loading factor in the turbulent-pressure-supported regime, $\etaturb$,
versus the disc gas fraction, $\fg$, for different
galaxy circular velocity ($\vc$) values (see the legend). For fixed $\vc$, $\etaturb$ decreases exponentially as $\fg$ decreases
below a critical value of $\sim 0.3$ because of the exponential cutoff in the outflow fraction, $\fw$ (\fref{fig:fw_vs_qfg}).
This cutoff implies that once a galaxy's gas fraction decreases below the critical value of $\sim 0.3$, it will no longer exhibit strong outflows.
The reason for this cutoff is that for Toomre-stable discs, $\sigt/\vc \propto \fg$.
Thus, as the gas fraction decreases, the turbulence becomes weak, the ISM becomes relatively smooth, and a negligible fraction of
the ISM is contained in sub-patches that can be blown out (i.e. that have $\siggas < \siggasmax$). (N.B. The red line is truncated
at the $\fg$ value below which it would no longer be turbulent pressure-supported.)}
\label{fig:eta_vs_fg}
\end{figure}

We will focus on the dependences of $\etaturb$ on the product $\fg \vc$. \fref{fig:eta_vs_vc} (\ref{fig:eta_vs_fg})
illustrates the dependence of $\etaturb$ on $\vc$ ($\fg$) for fixed values of $\fg$ ($\vc$).
For sufficiently high values of $\fg$ and $\vc$, $\fwturb$ (not shown)
is approximately constant, with $\fwturb \sim 0.5$. Thus, of order half the momentum from stellar feedback
is injected into underdense regions, which are then ejected from the disc. The other half is injected into overdense regions,
which are not ejected but rather `cycle' over the disc scaleheight and contribute to the turbulent self-regulation of star formation.
Because $\fwturb$ is approximately constant for high values of $\fg$ and $\vc$, the predicted scalings for $\fg = 1$ are similar to
the $\eta \propto \vc^{-1}$ scaling expected for a momentum-driven
outflow \citep[e.g.][]{Murray2005}; the predicted scalings are slightly steeper than this because $\fw$ decreases mildly with $\vc$ even when
$\fg = 1$. As $\vc$ or $\fg$ decrease below a critical value, $\fwturb$ decreases exponentially, which causes $\etaturb$ to
be significantly less than that expected from the $\eta \propto \vc^{-1}$ scaling.

\fref{fig:eta_vs_fg} is particularly useful because it clearly illustrates the transition that occurs at the critical gas fraction of $\sim 0.3$. Above
this critical gas fraction, the mass-loading factor for fixed $\vc$ is approximately independent of gas fraction. However, as the gas fraction decreases
from $\sim 0.4$ to $\la 0.2$, the mass-loading factor decreases by multiple orders of magnitude.
This cutoff occurs for a few reasons:
first, as shown in \fref{fig:fw_vs_qfg}, for a given Mach number, $\fwturb$ decreases exponentially below some
$\qturb \fg$ value. Thus, as $\fg$ decreases, $\fwturb$ decreases. Furthermore, for fixed $\qturb$, equation (\ref{eq:sig_t_2}) implies
that $\sigt \propto \fg \vc$. Consequently, as $\fg$ or $\vc$ decreases, $\sigt$ and thus
(for fixed $T$) $\mathcal{M}$ decrease, and \fref{fig:fw_vs_qfg} shows that the value of $\fwturb$ at fixed $\qturb \fg$ decreases as $\mathcal{M}$
decreases because the gas surface density PDF becomes narrower. We stress that this cutoff is not captured in standard sub-resolution and
semi-analytic models for galactic outflows and, as we shall demonstrate below, it has important implications for galaxy evolution.

\section{Thermal pressure-supported regime} \label{S:therm}

We now consider the limit in which turbulence is weak and thus thermal pressure supports the disc
against gravitational collapse. In this limit, photo-heating dominates the gas heating \citep{Krumholz2009,Ostriker2010}.
Thus, to be in this regime, the gas must not be self-shielding because
once the gas is self-shielding, the heating rate will drop dramatically and the gas will collapse until it becomes
turbulent-pressure-supported. For the gas to be self-shielding,
\begin{equation} \label{eq:selfshield}
\siggas \ga 10 \msun ~\mathrm{pc}^{-2} \left(\frac{Z}{Z_{\odot}}\right)^{-1} \equiv \sigss
\end{equation}
\citep{KMT09II}.
We also require that $Q > 1$ for similar reasons. In this regime, $\cs \gg \sigt$. Thus,
\begin{equation} \label{eq:q_wt}
Q \approx \frac{\sqrt{2} \cs \Omega}{\pi G \langle\siggas\rangle} \equiv \qtherm.
\end{equation}
The requirement that $Q \approx \qtherm > 1$ implies
\begin{equation}
\langle\siggas\rangle < \frac{\sqrt{2} \cs \Omega}{\pi G} \equiv \Sigma_{Q = 1}.
\end{equation}
Assuming $T = 10^4 $K and thus $\cs \approx 12$ km s$^{-1}$,
\begin{equation} \label{eq:q_eq_1}
\Sigma_{Q=1} = 12 \left( \frac{\Omega}{10 ~{\rm Gyr}^{-1}} \right) \msun ~{\rm pc}^{-2}.
\end{equation}
Consequently, we consider a galaxy to be in the thermal-pressure-dominated regime if
\begin{equation} \label{eq:thermal_condition}
\langle \siggas \rangle < \min\left(\sigss,\Sigma_{Q=1}\right).
\end{equation}
Because $\Omega \sim 10$ Gyr$^{-1}$ (see \aref{S:galaxy_scaling_relations}),
the $\mathcal{Q} > 1$ criterion is almost always the stricter criterion (i.e. unless
$Z \ga Z_{\odot}$, which is generally not the case for low surface-density galaxies).

Below, we will focus on galaxy-averaged
properties when investigating how the mass-loading factor depends on stellar mass and
redshift. The empirically based scaling relations presented in \aref{S:galaxy_scaling_relations} indicate that few
galaxies are in this regime in a galaxy-averaged sense (\aref{S:regimes}). Thus, we do not address this limit in
detail in the main text. However, our theory can also be applied to galaxies in a resolved
manner, and the outskirts of galaxies can be in the thermal-pressure-supported regime.
For this reason, we derive the equilibrium SFR surface density relation and mass loading
factor in \aref{S:therm_app}, and we only quote the results here.
The equilibrium SFR surface density relation in this regime is
\begin{equation} \label{eq:sigsfr_wt_main}
\sigsfrtherm = \Omega Z \langle\siggas\rangle \left(\frac{\langle\siggas\rangle}{\Sigma_0}\right),
\end{equation}
where
\begin{equation}
\Sigma_0 \approx 3 \msun ~{\rm pc}^{-2}.
\end{equation}
The equilibrium SFR surface density relation given by equation (\ref{eq:sigsfr_wt_main}) is the equivalent of
equation (\ref{eq:cafg_eq18}) for galaxies in the thermal-pressure-supported regime.
We note that in both regimes, $\sigsfr \propto \msiggas^2$, but in the thermal-pressure-supported
regime, there is an additional dependence on the product $\Omega Z$. The mass-loading factor is
\begin{equation} \label{eq:eta_therm_main}
\etatherm = 15 \fw \left( \frac{Z}{Z_{\odot}} \right)^{-1} \left(\frac{\langle\siggas\rangle}{10 \msun ~{\rm pc}^{-2}}\right)^{-1},
\end{equation}
where $\fwtherm \sim 1$ for $Z\msiggas/\vc \ga 10^{-2} \msun$ pc$^{-2}$ km$^{-1}$ s and is exponentially suppressed
for $Z\msiggas/\vc \la 2 \times 10^{-3} \msun$ pc$^{-2}$ km$^{-1}$ s (\fref{fig:fw_thermal_vs_smovc}). Equation
(\ref{eq:eta_therm_main}) can be approximated as
\begin{eqnarray} \label{eq:eta_therm_approx}
\etatherm &\approx& 15 \left( \frac{Z}{Z_{\odot}} \right)^{-1} \left(\frac{\langle\siggas\rangle}{10 \msun ~{\rm pc}^{-2}}\right)^{-1} \\
&\times& \exp \left(\frac{-\vc/Z \msiggas}{500 ~\mathrm{km} ~\mathrm{s}^{-1} ~\msun^{-1} ~\mathrm{pc}^{2}}\right).\nonumber
\end{eqnarray}

\begin{figure}
\centering
\includegraphics[width=0.9\columnwidth]{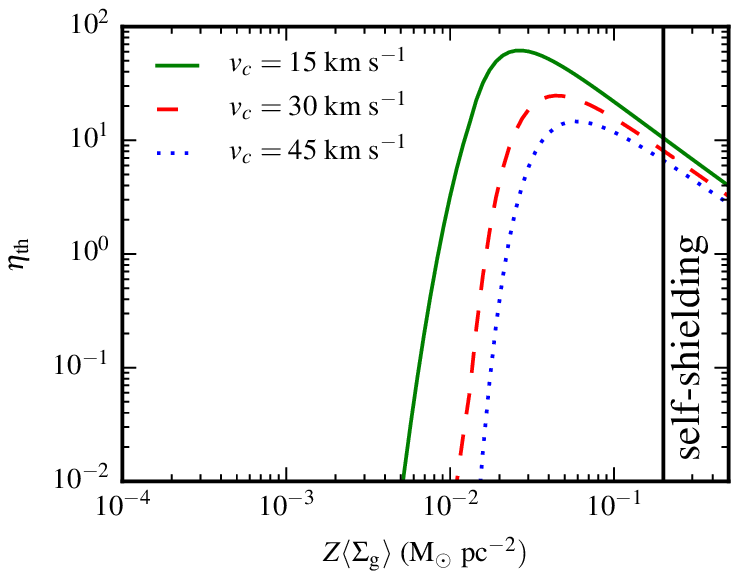}
\caption{The mass-loading factor in the thermal-pressure-supported regime, $\etatherm$, versus the mean metal surface density
$Z \msiggas$ for some relevant circular velocity values (see the legend).
The vertical black line denotes the self-shielding threshold; thermal-pressure-dominated galaxies lie to the left
of this line by definition. Additionally, they must be Toomre-stable, which implies $\msiggas < 12 (\Omega/\mathrm{Gyr}^{-1}) \msun$
pc$^{-2}.$ For such galaxies, $\eta$ can be significant if the metal surface density
is $\ga 10^{-2} \msun$ pc$^{-2}$, i.e. within one order of magnitude of the self-shielding threshold.
The mass-loading factor is suppressed at lower surface densities because the momentum deposition rate per area that
corresponds to the SFR surface density required to maintain self-regulation via photo-heating is too low to accelerate a significant fraction
of the ISM to the escape velocity on a coherence time.}
\label{fig:fw_thermal_vs_sig_met}
\end{figure}

\fref{fig:fw_thermal_vs_sig_met} shows $\etatherm$ versus $Z \msiggas$ for $\vc = 15$, 30
and $45 \kmpers$. The black vertical line indicates the self-shielding threshold;
for a patch to be in the thermal-pressure-dominated regime, it must have $Z \msiggas$ less than this limit. When a
galaxy/patch is near the self-shielding limit, $\fwtherm \sim 0.5$, whereas for $Z \msiggas \la 10^{-2} \msun$ pc$^{-2}$,
$\fwtherm \la 10^{-3}$. This indicates that there is a `sweet spot', $Z \msiggas \sim 0.1 - 1 \sigss$,
in which a galaxy can both be supported by thermal pressure and exhibit strong outflows. Because $\etatherm \propto
\fwtherm (Z\msiggas)^{-1}$, $\etatherm$ is maximal for patches with $Z \msiggas \sim 0.1 \sigss$.

\section{Dependence of the mass-loading factor on stellar mass and redshift} \label{S:eta_global_props}

\begin{figure}
\centering
\includegraphics[width=0.9\columnwidth]{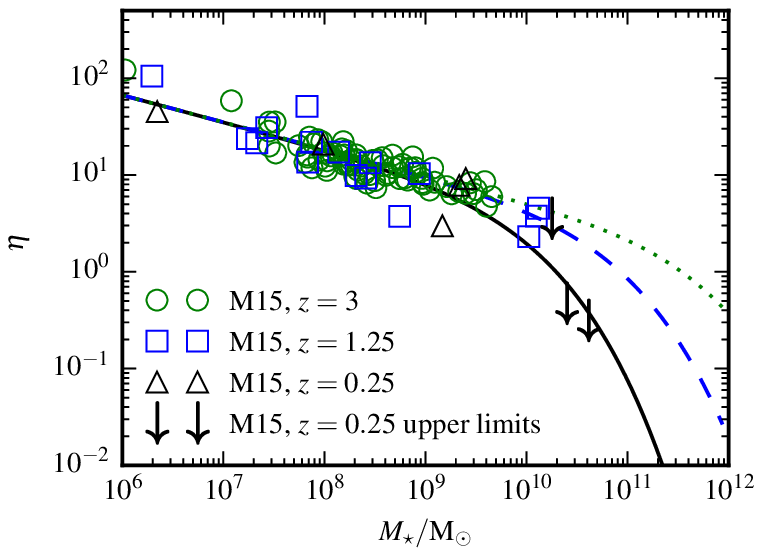}
\caption{The mass-loading factor, $\eta$, versus stellar mass, $\mstar$, at $z = 0.25$ (solid black), 1.25 (dashed blue),
and 3 (dotted green). The mass-loading factor values for
the FIRE simulations \citep{Muratov2015} are indicated by the open symbols; the arrows denote upper limits for
massive galaxies at $z = 0.25$.
We have assumed that the order-unity parameter $\phi/\mathcal{F}'$ in equation (\ref{eq:sigmax_turb_final}) is
0.7, as this yields the best agreement with the simulation results. In massive galaxies, outflows are significantly
suppressed at low redshift because their gas fractions
decrease below the critical value at which the outflow fraction decreases exponentially (see \fref{fig:eta_vs_fg}).
The reason for this cutoff is that the turbulent velocity dispersion, $\sigt$, required to provide vertical pressure support
decreases with the gas fraction. When $\sigt/\vc$ becomes too low, the amount of mass in sufficiently underdense patches
becomes negligible.
In contrast, outflows are driven effectively in low-mass galaxies ($\mstar \la 10^{10} \msun$) at all redshifts.}
\label{fig:eta_vs_mstar}
\end{figure}

For comparison with simulations and observations, we wish to investigate how the mass-loading factor predicted by our model depends
on stellar mass and redshift. To do so, we need to know how the disc-averaged gas fraction and metallicity, the effective radius,
and the circular velocity and orbital frequency at the effective radius scale with galaxy mass and redshift.
It is not our purpose here to predict these quantities a priori. Thus, we will for now simply adopt empirically based fitting
functions to interpolate between the observations at different redshifts. The relevant fitting functions for $\fg$, $\vc$, $Z$, and
$\re$, from which scaling relations for $\Omega$, $\msiggas$ and $\sigt$ are derived,
are presented and plotted as a function of galaxy mass and redshift for reference in \aref{S:galaxy_scaling_relations}.
Because we rely on various empirically based scaling relations, the $\eta(\mstar,z)$ relation that we present is by no means an
ab initio prediction of our model. For this reason, those interested in implementing our model in simulations or SAMs
should ideally not use the $\eta(\mstar,z)$ relation but rather the physical scalings presented above (see \sref{S:implementation}
for details); nevertheless, we will provide a fitting function for $\eta(\mstar,z)$ below.

Our assumed scaling relations imply that based on their global properties, all galaxies are expected
to be turbulent-pressure-supported (see \aref{S:regimes}). Only galaxies that have $\msiggas \la 10 \msun$ pc$^{-2}$, and thus lie
below the empirically based $\msiggas(\mstar,z)$ relation plotted in \fref{fig:sig_gas_vs_mstar}, can be globally supported by
thermal pressure. We thus explore how the mass loading
factor depends on stellar mass under the assumption that all galaxies are in the turbulent-pressure-supported regime.
By combining equations (\ref{eq:f_w}), (\ref{eq:M}), (\ref{eq:eta_fg_vc}), (\ref{eq:f_gas}), and (\ref{eq:TF}) and
assuming an ISM temperature $T$, we can calculate the dependence of $\fw$ and $\eta$ on $\mstar$ and $z$.
The resulting relations are presented in \fref{fig:eta_vs_mstar}, which shows
$\eta$ versus $\mstar$ for $z = 0.25$, 1.25 and 3.
We overplot the mass-loading factors for the FIRE
simulations \citep{Muratov2015} at the same redshifts as our model predictions. Note that the black arrows
denote upper limits: as discussed by \citet{Muratov2015}, galaxies in the FIRE simulations with $\mstar \ga 10^{10} \msun$
exhibit negligible mass outflow rates at $z \la 1$. Note that for this plot only, we assume that the order-unity parameter
$\phi/\mathcal{F}' = 0.7$ because of the few near-unity values that we tried, this gave the best agreement with the
simulation results.

At high redshift, outflows are driven effectively for all $\mstar$ values. 
At $z = 3$, the mass-loading factor varies from $\sim 70$ at $\mstar = 10^6 \msun$ to $\sim 0.5$ at
$\mstar = 10^{12} \msun$. For high-mass ($\mstar \ga 10^{10} \msun$) galaxies, as the redshift decreases, $\fw$
and thus $\eta$ decrease significantly.
For $\mstar \sim 10^{11} \msun$, $\eta$ decreases by almost two orders of magnitude from $z = 3$ to $z = 0.25$.
This decrease in $\eta$ occurs because the gas fractions of massive galaxies decrease below the critical value at which $\fw$ decreases
exponentially (see \fref{fig:eta_vs_fg}). In contrast, in low-mass ($\mstar \la 10^9 \msun$) galaxies, $\eta$
is high and approximately redshift-independent at all redshifts because these galaxies are sufficiently gas-rich
even at $z = 0$.

For galaxies with $\mstar \la 10^{10} \msun$, the redshift evolution in the $\eta$ -- $\mstar$ relation is
weak. The reason is as follows: recall that $\etaturb \propto \fw (\fg \vc)^{-1}$. 
The galaxy circular velocity (unlike the halo circular velocity) at fixed $\mstar$ is independent of redshift.
The outflow fraction depends only weakly on the gas fraction as long as $\fg \ga 0.3$ (which holds
for these galaxies according to our assumed empirical relation), and the decrease in $\fw$ with
decreasing $\fg$ is approximately canceled by the $\fg^{-1}$ term. Thus, for galaxies with $\mstar \la 10^{10} \msun$,
$\eta$ scales as $\sim \vc^{-1}$ with a normalization that is approximately independent of redshift.

The agreement between our predictions and the simulation results of \citet{Muratov2015} is impressive, especially
given the simplicity of our model. In particular, a novel feature of our model is that it explains the suppression of
outflows in massive galaxies at low redshift. This cutoff occurs because at low redshift, the gas fraction and thus
mean gas surface density of massive galaxies decreases considerably. Consequently, the turbulent velocity
expected from self-regulation decreases, and thus both $\siggasmax$ and the Mach number decrease. The fraction
of the ISM that can be accelerated to the escape velocity on a dynamical time becomes negligible, and outflows
are suppressed. We discuss the implications of this suppression below.

\section{Implications for galaxy evolution} \label{S:implications}

We have demonstrated that our model predicts that the fiducial scaling $\eta \propto \vc^{-1}$ will not hold for all star-forming galaxies
because in our model, $\eta \propto \fw (Q \fg \vc)^{-1} \propto \fw \sigt^{-1}$. Thus, even if we assume that $Q = 1$ (which,
as we discuss below, is a simplification), the $\eta \propto \vc^{-1}$ scaling can be altered by variations in $\fg$ (or equivalently $\sigt$)
and $\fw$ (which is determined by the values of $Q$, $\fg$, $\vc$, and the diffuse ISM temperature). Consequently, models
that assume a fixed $\eta \propto \vc^{-1}$ relation (or, more generally, any model that assumes that $\eta$ depends on a single
global property, such as $\vc$) may miss some important physical effects and artificially reduce the dispersion in
galaxy properties because variations in $\fg$ at fixed $\vc$ will directly lead to variations in $\eta$ at fixed $\vc$.

For example, suppose that a galaxy's gas fraction is temporarily lower than the quasi-equilibrium value. This could occur because of
the stochastic nature of gas accretion from the intergalactic medium \citep[e.g.][]{KeresHernquist2009} or because the galaxy
is in a post-burst state and its gas fraction has not returned to the steady-state value set by the balance
of inflow and outflow. In standard models, the galaxy's mass-loading factor would remain approximately constant
(assuming that $\vc$ is approximately constant),
so its outflow rate would decrease by the same factor as its SFR. Thus, the galaxy's gas content would continue to be depleted
by star formation and outflows at a rate of $(1+\eta) \sfr$ and would need an inflow rate greater than this value in order for its gas
fraction to become high again. In contrast, in our model, if the gas fraction becomes sufficiently low ($\fg \la 0.3$),
the mass-loading factor is exponentially suppressed, and the depletion rate becomes equal to the SFR alone. Thus, as long
as the inflow rate is greater than the SFR, the gas fraction will increase until the galaxy re-enters the high-outflow regime.
The difference between these two depletion/required infall rates can be dramatic in $\mstar \la 10^{10} \msun$ galaxies, for which $\eta \gg 1$.
This example illustrates that assuming a pure $\eta \propto \vc^{-1}$ scaling without accounting for the cutoff that we demonstrate
can artificially prevent low-mass galaxies from accumulating fresh gas and artificially suppress the variability of star formation.

The behavior can also differ in periods in which the gas surface density has been driven higher than the equilibrium value, e.g.
because of an interaction. As the ISM is compressed, the SFR and thus outflow rate will increase. Assuming $\eta \propto \vc^{-1}$
and $\sigsfr \propto \msiggas^2$, for fixed $\vc$, the mass outflow rate surface density scales as $\sigwind \propto \msiggas^2$.
However, our model predicts that in both regimes, all else being equal, $\eta \propto \fw \msiggas^{-1}$.
$\fw$ increases with $\msiggas$, but this increase is weak if the galaxy already has high $\fw$ (i.e. if $\siggasmax$ is
not in the low-$\siggas$ tail of the surface density PDF). Suppose that this is the case and thus $\fw$ is approximately constant
(otherwise, arguments similar to those in the previous paragraph apply).
Then, $\sigwind \propto \msiggas ^{-1}$. Consequently, models that assume fixed $\eta \propto \vc^{-1}$ scalings will tend to have
greater outflow rates in starbursts than predicted by our model. These stronger outflow rates would result in artificial suppression
of starbursts in such models. This effect may partially explain why large-volume cosmological simulations such as Illustris
\citep{Vogelsberger2014} do not exhibit strong starbursts \citep{Sparre2015}, whereas galaxies in the FIRE simulations have SFRs that can vary by multiple
orders of magnitude on 100-Myr timescales \citep{Hopkins2014,Muratov2015,Sparre2015fire}, although resolution
is another crucial difference between these two types of simulations. We note that suppression of starbursts is especially undesirable
because they are critical for making realistic bulges \citep{Hopkins:2008extra_light,Hopkins:2009cusps,Hopkins:2009cores}.

Moreover, as mentioned above, our model predicts that massive galaxies transition from being highly turbulent, clumpy discs that exhibit
strong outflows at high redshift to well-ordered discs characterised by weak or non-existent outflows at low redshift
because their gas fractions decrease below a critical value of $\sim 0.3$. Suppose that
the product $Q \fg$ (or equivalently $\sigt/\vc$) decreases below the critical value such that $\fw$ is exponentially suppressed.
The subsequent evolution depends critically on the gas inflow rate:
if the gas inflow rate is greater than the SFR, the galaxy will accumulate gas (because $\eta \sim 0$), $Q \fg$ will increase above
the critical value, the galaxy will reenter the quasi-equilibrium state, and $\eta$ will return to the equilibrium value.
Thus, at high redshift, when gas inflow rates are high, the quasi-equilibrium can be
maintained, and galaxy evolution should be characterized by bursts of star formation (because of stochasticity in the inflow rate and
mergers) and strong outflows. However, if the gas supply is insufficient to drive $Q \fg$ back above the critical value, then the galaxy will enter a
steady mode of star formation in which $\eta \sim 0$ and $\sigt/\vc \ll 1$. The results of \citet{Muratov2015} indicate that galaxies
with $\mstar \ga 10^{10} \msun$ transition to this regime by $z = 0$, whereas lower-mass galaxies remain in the high-$\eta$ regime.
This transition is potentially critical for the ability to form a thin, cold molecular (and hence young stellar) disc.
In many SAMs and cosmological simulations, it is challenging to suppress early star formation, as is necessary to form disc
galaxies without overly massive bulges, without destroying discs and over-suppressing star formation at $z \sim 0$
\cite[e.g.][]{Torrey2014,White2015}. The outflow model that we present here provides a natural mechanism to address this problem,
and the results of the FIRE simulations support this explanation.

We have argued that galaxies self-regulate to maintain a quasi-equilibrium state in which $Q \sim 1$ globally. The reader may object
that this is inconsistent with observations of `clumpy' galaxies at $z \sim 2$ \citep[e.g.][]{Genzel2011}. However, we stress that
this does not mean that all regions of a galaxy have $Q \sim 1$, and in fact, our model relies on overdense sub-patches of a galaxy having
$Q < 1$; these patches would correspond to the observed massive star-forming clumps.
Patches with $Q < 1$ collapse, form stars, and inject momentum into the surrounding ISM, thereby maintaining a global $Q$ of
order unity. Even if it is possible for galaxies to have $Q < 1$ globally, and thus be (temporarily) globally unstable, our model can be applied
because we have retained the $Q$ dependences in all expressions. Because $\fw$ depends on the product $Q \fg$, a galaxy with global
$Q = Q_1 < 1$ and gas fraction $\fg = f_{\mathrm{g,1}}$ will have the same $\fw$ as a galaxy with $Q = 1$ and gas fraction
$\fg = Q_1 f_{\mathrm{g,1}}$. For fixed $\fw$, $\fg$, and $\vc$, $\eta \propto Q^{-1}$. If $Q \fg$ is sufficiently high that $\fw$ is approximately
constant, a galaxy with global $Q < 1$ would have a greater outflow rates than a stable galaxy with the same $\fg$ and $\vc$ values, which
would tend to drive the unstable galaxy back toward stability. However, if the global $Q$ is low enough that $Q \fg$ is less than the critical
value at which $\fw$ is exponentially cutoff, outflows will be suppressed, and the galaxy will unstably fragment until the gas surface density
is reduced by star formation to the value required to achieve $Q = 1$. Whether the galaxy will again exhibit strong outflows will depend
on the subsequent evolution of $Q \fg$.

It is clear that our model predictions should differ from those of the types of models currently employed in large-volume cosmological
simulations and SAMs. Thus, it would be very interesting to implement our model in a cosmological simulation or SAM to investigate
e.g. the implications of the cutoff discussed above. For this reason, we will now outline how one could do so.

\section{How to implement the model in simulations or semi-analytic models} \label{S:implementation}

As discussed above, our model can be applied to both entire galaxies and to sub-regions of galaxies, provided that the
size of the sub-region is greater than the disc scaleheight.
Given that state-of-the-art cosmological simulations do not
resolve disc scaleheights, it would be natural to apply our model to individual resolution elements in such
simulations. In SAMs, one could apply it either to individual galaxies or in annuli; the latter approach would capture potentially
important effects such as variations in $\fg$ with radius that were ignored in the above analysis. To calculate the mass loading
factor for a given galaxy, resolution element or annulus, one should proceed as follows.

\begin{enumerate}
\item Depending on whether the entire galaxy or a region (i.e. resolution element or annulus) is considered, use the global or
local values for $\Omega$ and $Z$ to calculate the global or local values for $\Sigma_{Q=1}$ and
$\sigss$ using equations (\ref{eq:q_eq_1}) and (\ref{eq:selfshield}), respectively.
\item If the value of $\msiggas$ is less than both the $\Sigma_{Q=1}$ and $\sigss$ values, the galaxy/resolution
element/annulus is in the thermal-pressure-supported regime. The mass-loading factor can then be calculated using
equations (\ref{eq:xw_thermal}), (\ref{eq:mach_thermal}), (\ref{eq:f_w}), (\ref{eq:tk14_eq12}), (\ref{eq:tk14_eq14}) and
(\ref{eq:eta_therm_main}). Alternatively, the approximate relation given in equation (\ref{eq:eta_therm_approx}) can be employed.
\item Otherwise, the galaxy/resolution element/annulus is supported by turbulent pressure.
Equations (\ref{eq:xw_turb}), (\ref{eq:M}), (\ref{eq:f_w}), (\ref{eq:tk14_eq12}), (\ref{eq:tk14_eq14}) and (\ref{eq:eta_fg_vc})
can then be used to calculate the mass-loading factor.
\end{enumerate}

Although we advocate using the method presented above for calculating the mass-loading factor, the scalings of
$\eta$ with the products $\fg \vc$ and $\fg \mstar$, including the cutoff, can be
approximately captured using the following fitting functions:
\begin{eqnarray}
\eta &\approx& 15 \left(\frac{\fg \vc}{100 ~\kmpers}\right)^{-1} \exp \left(\frac{-0.75}{\fg}\right) \label{eq:eta_fgvc_approx} \\
&\approx& 14 \left(\frac{\fg \mstar}{10^{10} ~\msun}\right)^{-0.23} \exp \left(\frac{-0.75}{\fg}\right). \label{eq:eta_fgmstar_approx}
\end{eqnarray}

\section{Conclusions} \label{S:conclusions}

We have presented an analytic model for how stellar feedback simultaneously regulates star formation and drives outflows
in a turbulent ISM. The model is based on two fundamental assumptions: (1) star formation is self-regulating: at high surface density,
the momentum input from star formation drives turbulence and supports the disc against gravitational collapse;
at low surface density, when turbulence is subdominant, photo-heating stabilizes the disc.
(2) Because of turbulence, the ISM exhibits a range of gas surface densities. There is a critical surface density below
which gas can be accelerated to the escape velocity before the density field is `reset' by turbulence.
The fraction of the ISM with surface density less than this critical value is expelled in an outflow.

Our principal conclusions are as follows:
\begin{enumerate}
\item In most galaxies (i.e. those with $\mstar \ga 10^6$ at $z = 0$, and even less massive galaxies at higher redshift),
within the effective radius, turbulent pressure dominates the pressure
support against gravity. For such galaxies,
if the gas fraction is above the critical value of $\sim 0.3$, a few tens of percent of the ISM is expelled as an outflow
over an orbital time, and the outflow mass-loading factor $\eta \equiv \dot{M}_{\rm out}/\sfr$ scales as $\eta \propto (\fg \vc)^{-1}$,
where $\fg$ is the gas fraction of the disc and $\vc$ is the \emph{local} circular velocity. Because
at fixed redshift, galaxy gas fractions decrease with stellar mass, this scaling is slightly steeper than that
expected for a momentum-driven outflow, $\eta \propto \vc^{-1}$. When the gas fraction (or equivalently $\sqrt{2} \sigt/\vc$, where
$\sigt$ is the turbulent velocity dispersion) is less than the critical value of $\sim 0.3$, $\eta$
is significantly less than expected from the $\eta \propto (\fg \vc)^{-1}$ scaling. The reason for this suppression
is that when the turbulent velocity dispersion is low relative to the circular velocity, only patches with surface densities
significantly less than the mean can be blown out. However, because of the low $\sigt/\vc$, the ISM is relatively
smooth, and a negligible fraction of the ISM mass is contained in patches with surface density less than the critical
value. The cutoff can be approximated using the fitting functions that we present in equations (\ref{eq:eta_fgvc_approx})
and (\ref{eq:eta_fgmstar_approx}).
\item In the low-$\siggas$ outskirts of galaxies and in galaxies that have unusually low $\msiggas$, thermal pressure
provides the dominant support against gravity. In this regime, photo-heating from star formation rather than
momentum deposition from stellar feedback maintains equilibrium. To be in this regime, a galaxy or patch of ISM must
both be non-self-shielding and have $Q > 1$. Outflows can be driven effectively when the mean metal surface density
is at least one-tenth the self-shielding limit, with $\eta \propto (Z \msiggas)^{-1}$. Below this value, the mass-loading factor
is exponentially suppressed.
\item Because low-mass galaxies are relatively gas-rich (and thus turbulent, assuming that they self-regulate to
$Q \sim 1$) at all redshifts, they should exhibit mass-loading factors that are high (e.g. $\eta \sim 20$ for
$\mstar \sim 10^8 \msun$) and approximately redshift-independent at fixed stellar mass. This redshift
independence occurs because the mass-loading factor depends on the galaxy circular velocity, not the halo circular
velocity. Thus, models that use the halo circular velocity to calculate $\eta$, as has been often done in SAMs and cosmological
simulations, will introduce redshift evolution in the $\eta$ -- $\mstar$ relation that is inconsistent with our theory.
\item In massive galaxies ($\mstar \ga 10^{10} \msun$), outflows are suppressed at low
redshift because the gas fraction, and thus turbulent velocity dispersion, of such galaxies decreases
to values ($\fg \la 0.3$) for which the outflow fraction decreases exponentially. For $\mstar \approx 10^{11} \msun$ galaxies,
the mass-loading factor decreases by almost two orders of magnitude from $z = 3$ to $z = 0$.
\item We predict that this suppression of outflows is necessary for the emergence of steadily star-forming, stable,
thin molecular gas (and thus young stellar) discs. Because the critical gas fraction is $\sim 0.3$, only moderately
massive, low-redshift galaxies can form such discs. Galaxies in which outflows are not suppressed should be characterised
by cycles of starbursts, outflows, and quenched periods. This prediction is supported by the results of the FIRE simulations
\citep{Hopkins2014,Muratov2015,Sparre2015fire}.
\item The mass-loading factors predicted by our model depend on additional properties besides $\vc$. Thus,
we expect galaxies to exhibit variations in $\eta$ at fixed $\vc$; such variations are excluded by construction
in the models currently assumed in large-volume cosmological simulations and SAMs.
\end{enumerate}

The simple analytic model presented here is attractive because it provides a potential physical explanation for
the behaviour of outflows observed in state-of-the-art simulations that explicitly model stellar feedback
\citep{Muratov2015}. However, one must confirm that the various assumptions in the model hold and that
further predictions of the model (e.g. weak turbulence in massive galaxies at low redshift) are borne out by
the simulations; this is one avenue for future work. Moreover, it will be of interest to incorporate the
predicted mass-loading factor scalings in cosmological simulations and SAMs.

\acknowledgments

We thank Lee Armus, Benham Darvish, Claude-Andr\'e Faucher-Gigu\`ere, Tim Heckman, Andrey Kravtsov,
Crystal Martin, Sasha Muratov, Eve Ostriker,
Joel Primack, and Rachel Somerville for useful discussions, Sasha Muratov for providing data
from \citet{Muratov2015} in electronic form,
and Andrey Kravtsov for noting a typo. \rev{We also thank the reviewer for a constructive report that helped improve the
quality of the manuscript.}
CCH is grateful to the Gordon and Betty Moore Foundation for financial support, and he is especially
grateful to Emmett Hayward for motivating rapid completion of the manuscript by his impending arrival, Tara
Hayward for enabling his arrival, and Lori Diebold for facilitating both the completion of the manuscript and the arrival of
Emmett. Support for PFH was provided by an Alfred P. Sloan Research Fellowship, NASA ATP Grant NNX14AH35G,
and NSF Collaborative Research Grant \#1411920 and CAREER grant \#1455342.
This work was supported in part by National Science Foundation Grant No. PHYS-1066293
and the hospitality of the Aspen Center for Physics, and it benefitted greatly from CCH's participation in
the 2015 ACP Summer Program ``The Physics of Accretion and Feedback in the Circum-Galactic Medium'',
the BIRS-CMO workshop ``Computing the Universe: At the Intersection of Computer Science and Cosmology'',
and the KITP program ``The Cold Universe.''
This research has made use of NASA's Astrophysics Data System Bibliographic Services.
\\

\footnotesize{
\bibliographystyle{mn2e}
\bibliography{winds}
}

\begin{appendix}

\section{Derivations for the thermal-pressure-dominated regime} \label{S:therm_app}

Here, we derive the equilibrium surface density
and mass-loading factor relations for the thermal-pressure-dominated regime (equations
\ref{eq:sigsfr_wt_main} and \ref{eq:eta_therm_main} in the main text).

\subsection{The equilibrium SFR surface density relation}

In this regime, thermal pressure supports the disc against gravity. Thus, the equilibrium SFR will
be set by the requirement that photo-heating from star formation balances cooling.\footnote{The
metagalactic background radiation can also photoheat the gas, but this component is expected to
be important only in the far outer disc regions of galaxies, where $\siggas \ll 1 \msun$ pc$^{2}$ (see \citealt{Ostriker2010}
for a detailed discussion).}
For photoionization, which should dominate the heating, the heating rate per unit area is
\begin{equation} \label{eq:heating_rate}
\frac{\dot{E}_{\mathrm{heat}}}{A} = \frac{\beta L}{A} = \beta \epsilon c^2 \sigsfr,
\end{equation}
where $\beta \sim 0.1$ is the fraction of the stellar luminosity that is emitted as ionizing photons \citep{Leitherer1999}
and $\epsilon \sim 4 \times 10^{-4}$ is the
ratio of the energy emitted over the life of a star to its rest-mass energy. For other forms of heating, only the prefactor
$\beta$ differs. The cooling rate per unit area is
\begin{equation} \label{eq:cooling_rate}
\frac{\dot{E}_{\mathrm{cool}}}{A} = \frac{\Lambda n_{\mathrm{e}} n_{\mathrm{i}} V}{A} \approx \Lambda \mu^{-1} \msiggas Z n_{\mathrm g},
\end{equation}
where $n_{\mathrm{e}}$, $n_{\mathrm{i}}$, and $n_{\mathrm{g}}$ are the electron, ion, and total gas number densities, respectively,
$V$ is the volume, $Z$ is the metallicity, and $\Lambda \sim 10^{-22}$ ergs s$^{-1}$ cm$^{3}$ is the net cooling rate
\citep[e.g.][]{Robertson2008}. Setting equations (\ref{eq:heating_rate}) and (\ref{eq:cooling_rate}) equal, we have
\begin{equation} \label{eq:sigsfr_phot_1}
\sigsfr = \frac{\Lambda}{\mu \beta \epsilon c^2} Z n_{\mathrm{g}} \langle \siggas \rangle \equiv \sigsfrtherm .
\end{equation}
Using $n_{\mathrm{g}} = \rho_{\mathrm{g}}/\mu$, $\rho = \langle \siggas \rangle/2 h$ (where $h$ is the disc scale height) and
$h = \sqrt{\cs^2 + \sigma_{\mathrm{T}}^2}/\sqrt{2}\Omega \approx \cs/\sqrt{2}\Omega$ (because we assume that thermal pressure
dominates the pressure support in this regime), we have
\begin{equation} \label{eq:n_g}
n_{\mathrm{g}} = \frac{\langle\siggas\rangle \Omega}{\sqrt{2} \mu \cs}.
\end{equation}
Combining equations (\ref{eq:sigsfr_phot_1}) and (\ref{eq:n_g}), it follows that
\begin{eqnarray}
\sigsfrtherm &=& \frac{\Lambda}{\sqrt{2} \mu^2 \beta \epsilon c^2 \cs} Z \langle\siggas\rangle^2 \Omega\\
&=& \Omega Z \langle\siggas\rangle \left(\frac{\langle\siggas\rangle}{\Sigma_0}\right), \label{eq:sigsfr_wt}
\end{eqnarray}
where we have defined
\begin{equation}
\Sigma_0 \equiv \frac{\sqrt{2} \mu^2 \beta \epsilon c^2 \cs}{\Lambda} \approx 3 \msun ~{\rm pc}^{-2},
\end{equation}
assuming $T = 10^4 $K and thus $\cs \approx 12$ km s$^{-1}$.
The equilibrium SFR surface density relation given by equation (\ref{eq:sigsfr_wt}) is the equivalent of
equation (\ref{eq:cafg_eq18}) for galaxies in the thermal-pressure-supported regime.
We note that in both regimes, $\sigsfr \propto \msiggas^2$, but in the thermal-pressure-supported
regime, there is an additional dependence on $Z$ (and also $\Omega$, but the variation in this
quantity with stellar mass and redshift is much less than that in $Z$).

\subsection{Mass-loading factor}

With the equilibrium SFR surface density relation in hand, we can now calculate the surface density below which gas
can be blown out by momentum deposition from stellar feedback on a coherence time. Again, as discussed in
\sref{S:f_w}, we stress that we are assuming that the outflows are momentum-driven in this regime because
the mass-loading factor for energy-driven outflows is low, our assumed $P_{\star}/m_{\star}$ value implicitly
includes the energy-conserving phase of SN remnants, and most material that is at sufficiently low surface density
that SN remnants can break out of the disc while still in the energy-conserving phase will also satisfy the criterion
for being blown out in a momentum-driven outflow.
Using equation (\ref{eq:siggas_max_1}), we have
\begin{equation}
\siggasmax = \frac{1}{\sqrt{2}} \left(\frac{P_{\star}}{m_{\star}}\right) \vc^{-1} Z \langle\siggas\rangle \left(\frac{\langle\siggas\rangle}{\Sigma_0}\right) \equiv
\siggasmaxtherm.
\end{equation}
Consequently,
\begin{eqnarray}
\xw &=& \ln \left[ \frac{1}{\sqrt{2}} \left(\frac{P_{\star}}{m_{\star}}\right) \vc^{-1} Z \left(\frac{\langle\siggas\rangle}{\Sigma_0}\right) \right]
\equiv \xwtherm \\
&=& \ln \left[ 14 \left(\frac{P_{\star}/m_{\star}}{3000 \kmpers}\right) \left(\frac{Z}{Z_{\odot}} \right) \right. \label{eq:xw_thermal} \\
&\times& \left. \left(\frac{\vc}{\kmpers}\right)^{-1} \left(\frac{\langle \siggas \rangle}{\msun ~{\rm pc}^{-2}}\right) \right], \nonumber
\end{eqnarray}
where we have used $Z_{\odot} = 0.02$.
Combining equations (\ref{eq:eta}) and (\ref{eq:sigsfr_wt}) yields
\begin{eqnarray}
\eta &=& \fw Z^{-1} \left(\frac{\langle\siggas\rangle}{\Sigma_0}\right)^{-1} \equiv \etatherm \\
&=& 15 \fw \left( \frac{Z}{Z_{\odot}} \right)^{-1} \left(\frac{\langle\siggas\rangle}{10 \msun ~{\rm pc}^{-2}}\right)^{-1}.
\end{eqnarray}
We note that $\eta$ is inversely proportional to the optical depth through the disc, $\tau \propto Z \langle\siggas\rangle$.

To solve for the outflow fraction, $\fw$, we need to determine the Mach number, $\mathcal{M}$. We assume that stellar feedback drives the
turbulence, although turbulent pressure does not support the disc against gravity in this regime. Then,
to calculate the turbulent velocity dispersion, we equate the rate at which momentum is injected by stellar feedback (equation
\ref{eq:p-sfr}) to the rate at which turbulence dissipates momentum, $\dot{\Sigma}_{\mathrm{P, disp}} \sim
\langle \siggas \rangle \sigt \Omega$. This yields
\begin{equation}
\sigt \sim \left(\frac{P_{\star}}{m_{\star}}\right) \frac{\sigsfr}{\langle \siggas \rangle \Omega}.
\end{equation}
Using the equilibrium SFR surface density relation for this regime (equation \ref{eq:sigsfr_wt}), we have
\begin{eqnarray}
\sigt &\sim& \left(\frac{P_{\star}}{m_{\star}}\right) \frac{Z \langle \siggas \rangle}{\Sigma_0} \\
&=& 20 ~\kmpers \left(\frac{P_{\star}/m_{\star}}{3000 \kmpers}\right) \left(\frac{Z}{0.1 Z_{\odot}}\right) \\
&\times& \left( \frac{\langle \siggas \rangle}{10 \msun ~{\rm pc}^{-2}}\right). \nonumber
\end{eqnarray}
Thus,
\begin{eqnarray}
\mathcal{M} &\sim& 1.7 \left(\frac{P_{\star}/m_{\star}}{3000 \kmpers}\right) \left(\frac{Z}{0.1 Z_{\odot}}\right)
\left( \frac{\langle \siggas \rangle}{10 \msun ~{\rm pc}^{-2}}\right) \label{eq:mach_thermal} \\
&\times& \left(\frac{\mu}{0.6}\right)^{1/2} \left(\frac{T}{10^4 ~\mathrm{K}}\right)^{-1/2} \equiv \mtherm. \nonumber
\end{eqnarray}

\begin{figure}
\centering
\includegraphics[width=0.9\columnwidth]{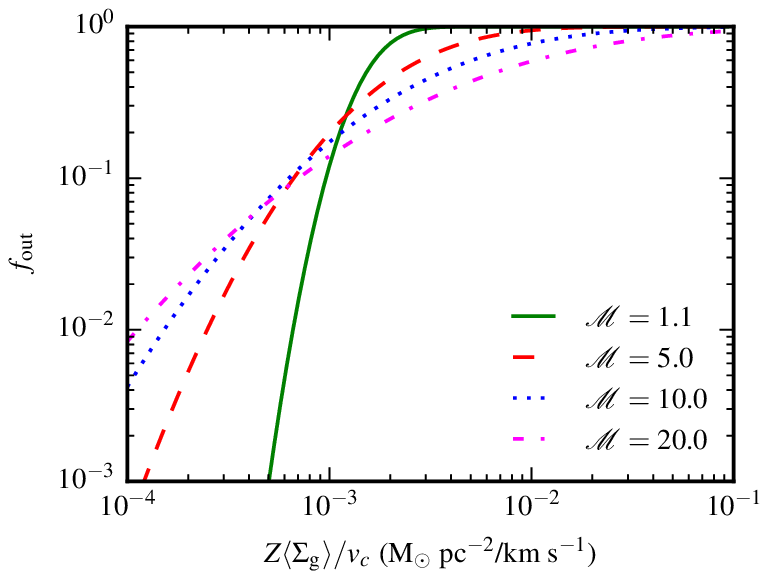}
\caption{The fraction of the ISM blown out per dynamical time in the thermal-pressure-dominated regime, $\fwtherm$,
versus $Z \langle \siggas \rangle/\vc$ for different Mach numbers (see the legend).
When $Z\msiggas/\vc$ (i.e. the mean metal surface density divided by the circular velocity) is $\ga 10^{-2} \msun$ pc$^{-2}$ km$^{-1}$ s,
$\fwtherm \sim 1$, which means that the entire ISM can be blown out on a dynamical time. At $Z\msiggas/\vc \la 10^{-3}
\msun$ pc$^{-2}$ km $^{-1}$ s, $\fwtherm$ is exponentially suppressed because the momentum deposition rate per area that
corresponds to the SFR surface density required to maintain self-regulation via photo-heating is too low to accelerate a significant fraction
of the ISM to the escape velocity on a coherence time.}
\label{fig:fw_thermal_vs_smovc}
\end{figure}

To calculate the outflow fraction, we perform the integral specified in equation (\ref{eq:f_w}) using $\xw =
\xwtherm$ (equation \ref{eq:xw_thermal}).\footnote{Note that when the turbulence becomes trans-sonic,
$\alpha = 3.7$ should be used in equation (\ref{eq:tk14_eq14}); see TK14 for discussion.}
\fref{fig:fw_thermal_vs_smovc} shows the outflow fraction $\fwtherm$ versus $Z \langle \siggas \rangle/\vc$ for Mach numbers
of 1.1, 5, 10, and 20. In all cases, $\fwtherm$ asymptotes to unity for $Z \langle \siggas \rangle/\vc \ga 10^{-2} \msun$ pc$^{-2}$ km$^{-1}$ s,
and it declines steeply below $\sim 10^{-3} \msun$ pc$^{-2}$ km$^{-1}$ s.

\section{Galaxy scaling relations} \label{S:galaxy_scaling_relations}

\begin{figure*}
\includegraphics[width=0.65\columnwidth]{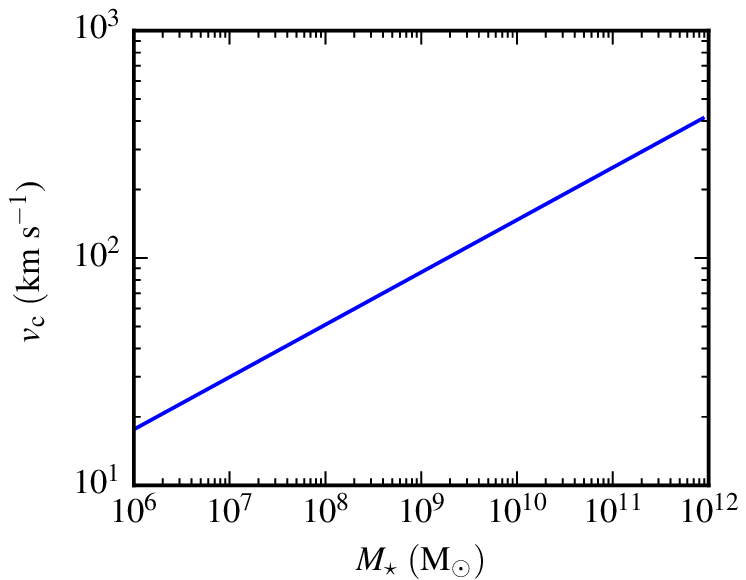}
\includegraphics[width=0.65\columnwidth]{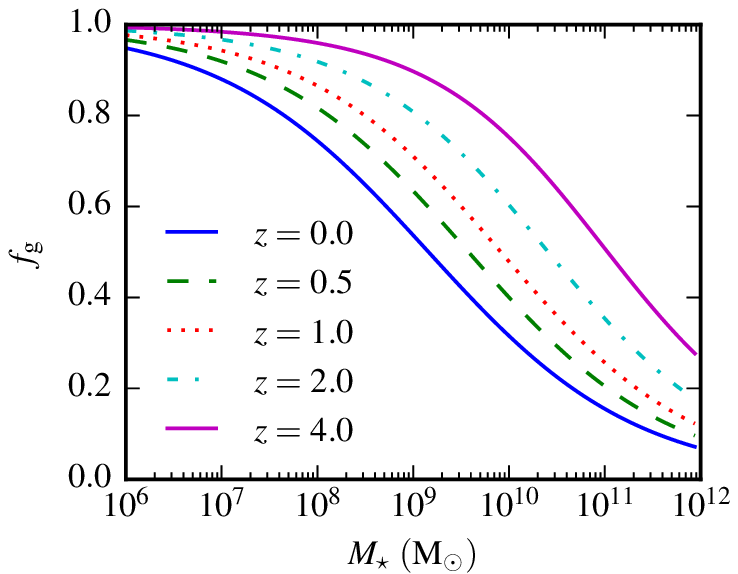}
\includegraphics[width=0.65\columnwidth]{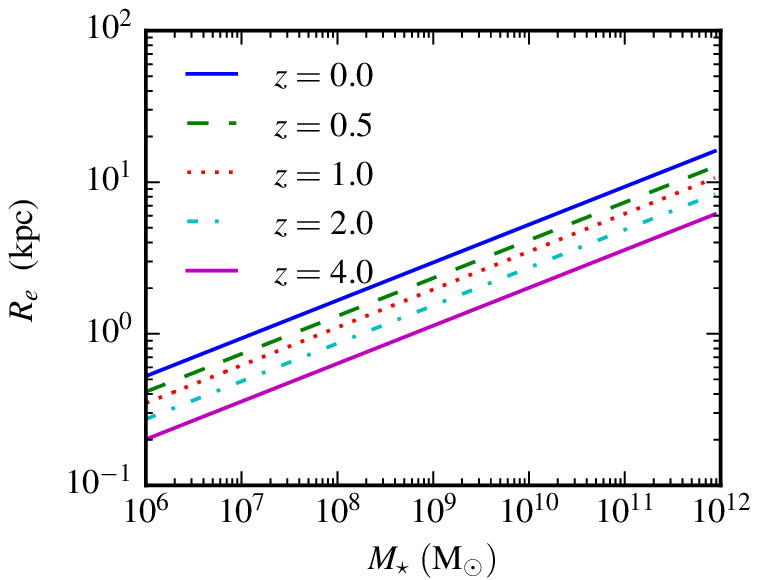}\\
\includegraphics[width=0.65\columnwidth]{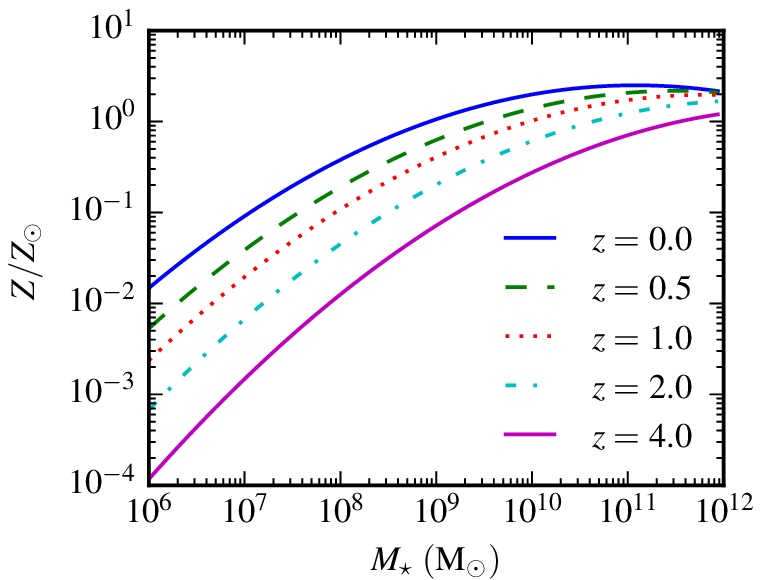}
\includegraphics[width=0.65\columnwidth]{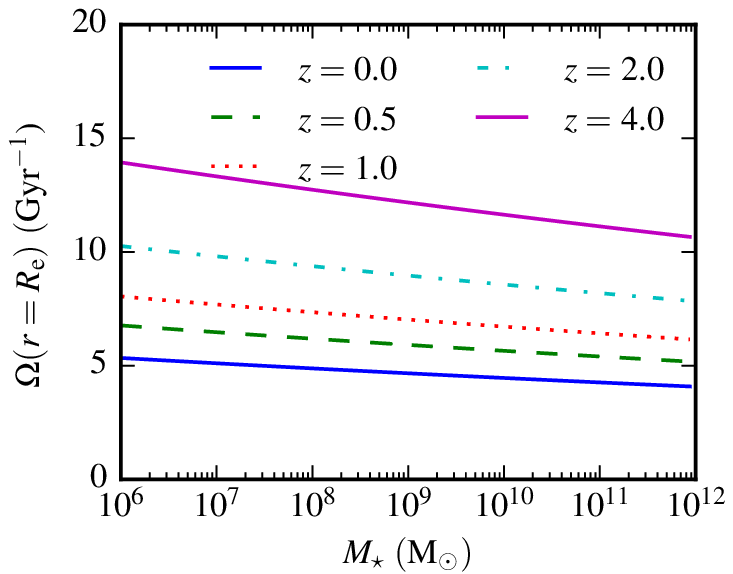}
\includegraphics[width=0.65\columnwidth]{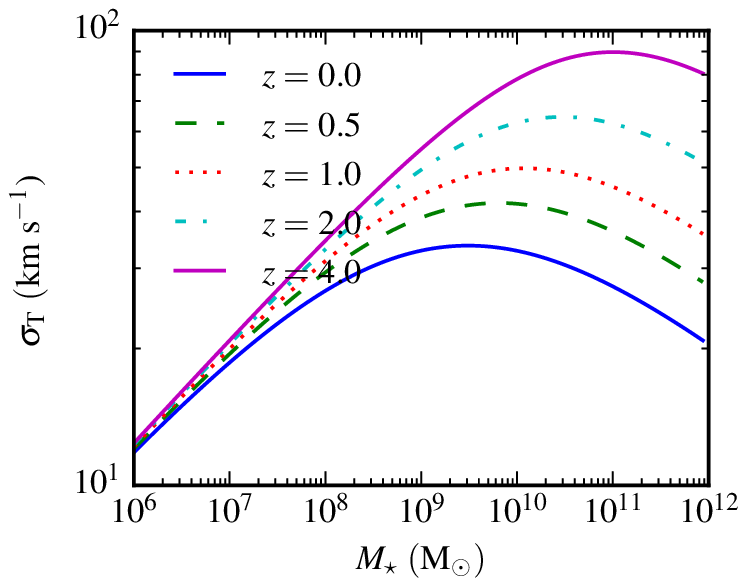}
\caption{Empirically based dependences of galaxy circular velocity (\emph{top left}), gas fraction (\emph{top middle}), effective radius (\emph{top right}),
metallicity (\emph{bottom left}), orbital frequency at the effective radius ($\Omega_{\re} \approx \vc/2\pi\re$; \emph{bottom middle}) and turbulent velocity
dispersion ($\sigt \approx \fg \vc/\sqrt{2}$ for $\qturb = 1$; \emph{bottom right}) on stellar mass
at various redshifts. The \emph{circular velocity} increases with stellar mass in a redshift-independent manner.
\emph{Gas fraction}: at fixed $z$, $\fg$ decreases with $\mstar$. At fixed $\mstar$, $\fg$ increases with $z$. \emph{Effective radius}
increases (decreases) weakly with stellar mass (redshift). \emph{Metallicity} increases (decreases) with stellar mass (redshift).
\emph{Orbital frequency}: for the redshift range $z = 0 - 4$, $\Omega_{\re}$ varies by less than a factor of 3 over 6 orders of magnitude in stellar mass.
It decreases (increases) very weakly with stellar mass (redshift). \emph{Turbulent velocity dispersion}: for $\mstar \la 10^8
\msun$, $\sigt \propto \vc$ because $\fg \sim 1$ at all redshifts. At higher stellar masses, the increase in $\vc$ with
stellar mass is counteracted by the decrease in $\fg$, thereby causing the value of $\sigt$ at fixed $z$ to decrease above
some $\mstar$ value. At fixed $\mstar$, $\sigt$ increases with $z$ because galaxies are more gas-rich at high redshift.}
\label{fig:scaling_relations}
\end{figure*}

To predict $\eta$ as a function of $\mstar$ and $z$, we require parameterizations
for the dependence of the circular velocity at the effective radius ($\vc$), the disc gas fraction [$\fg \equiv \mgas/(\mstar + \mgas)$],
the effective radius ($\re$) and the metallicity ($Z$) on $\mstar$ and $z$. For $\vc(\mstar)$, we assume that the stellar
\citet{TF} relation of \citet{BellDeJong2001},
\begin{equation} \label{eq:TF}
\vc = 147 \left(\frac{\mstar}{10^{10} \msun}\right)^{0.23} \kmpers,
\end{equation}
where we have divided the \citet{BellDeJong2001} masses by 1.7 to convert to a \citet{Kroupa2001} initial mass function, holds at all redshifts.
The relation is plotted in the upper-left panel of \fref{fig:scaling_relations}.
This specific (redshift-independent) parameterization is reasonable from $z = 0$ to at least $z \sim 1.7$
\citep{Miller2011,Miller2012,Miller2013}.

For $\fg$, we use the relation from \citet{Hopkins2009spheroids,Hopkins2010}:
\begin{eqnarray} \label{eq:f_gas}
\fg(\mstar | z = 0) & \equiv & f_0 \approx \left[1 + \left(\frac{\mstar}{10^{9.15} \msun}\right)^{0.4}\right]^{-1}, \nonumber \\
\fg(\mstar, z) & = & f_0 \left[1 - \tau(z)\left(1-f_0^{3/2}\right)\right]^{-2/3},
\end{eqnarray}
where $\tau(z)$ is the fractional lookback time to redshift $z$. This relation was obtained by fitting a theoretically
motivated functional form \citep{Keres2005,Keres2009} to a compilation of observations \citep{BellDeJong2000,
McGaugh2005,Shapley2005,Erb2006,Erb2008,Calura2008,Puech2008,Cresci2009,ForsterSchreiber2009,Mannucci2009}.
For convenience, we show $\fg$ versus $\mstar$ in the upper-middle panel of \fref{fig:scaling_relations}.

For the effective radius, we use the scaling relation from
\citet{Hopkins2010},
\begin{eqnarray}
\re(\mstar | z=0) &\approx& 5.28 ~\mathrm{kpc} \left(\frac{\mstar}{10^{10} \msun}\right)^{0.25}, \nonumber \\
\re(\mstar,z) &=& \re(\mstar | z = 0) (1+z)^{-0.6},
\end{eqnarray}
which is based on a compilation of observations \citep{Trujillo2004,Ravindranath2004,Ferguson2004,Barden2005,
Toft2007,Akiyama2008}. This relation is shown in the upper-right panel of \fref{fig:scaling_relations}.

For $Z$, we use the $Z(\mstar,z)$ relation presented in \citet{Hayward2013},\footnote{In
equation (12) of \citet{Hayward2013}, the exponent of the first $(1+z)$ term should be 0.094, not 0.94. The correct
value of 0.094 is stated in the text above the equation and was used in the calculations in that work.}
\begin{eqnarray}
\log \left(\frac{Z(\mstar,z)}{Z_{\odot}}\right) &=& -8.69 + 9.09(1+z)^{-0.017} \nonumber \\
&-& 0.0864 \left[ \log \left(\frac{\mstar}{\msun}\right) \right. \nonumber \\
&-& \left. 11.07 (1+z)^{0.094} \right]^2,\\
\end{eqnarray}
which is based on based on observations from \citet{Savaglio2005}, \citet{Erb2006MZ} and \citet{Maiolino2008}. For reference, the
relation is plotted in the bottom-left panel of \fref{fig:scaling_relations}.

It is interesting to consider some of the scalings that result from the above scalings. The bottom-middle panel of
\fref{fig:scaling_relations} shows the orbital frequency at the effective radius, $\Omega_{\re} \approx \vc/2\pi \re$, versus
$\mstar$ for various redshifts. At fixed $z$, $\re \propto \mstar^{0.25}$, and $\vc \propto \mstar^{0.26}$ at all redshifts. Thus,
at fixed $z$, $\Omega_{\re}$ is approximately constant, and it increases mildly from $\sim 5$ Gyr$^{-1}$ at $z = 0$ to $\sim 10-15$ Gyr$^{-1}$
at $z = 4$.

The bottom-right panel of \fref{fig:scaling_relations} shows
how the turbulent velocity dispersion (equation \ref{eq:sig_t_2}) depends on $\mstar$ at different redshifts; we assume $\qturb = 1$.
$\sigt$ ranges from $\sim 10 \kmpers$ at $\mstar \sim 10^6 \msun$ to $\sim 100 \kmpers$ in $\mstar \sim 10^{11} \msun$
discs at $z = 4$. For $\mstar \la 10^8 \msun$, $\fg \sim 1$ at all redshifts. Thus, $\sigt \approx \vc/\sqrt{2}$, which is related to the stellar
mass by the assumed Tully-Fisher relation. At higher stellar masses, $\sigt$ is approximately constant at
fixed redshift because the increase in $\vc$ with stellar mass is offset by the decrease in $\fg$. For fixed $\mstar \ga 10^8 \msun$,
$\sigt$ increases with $z$ because $\fg$ increases with $z$; i.e. high-redshift massive disc galaxies are much more turbulent
than their local counterparts.

\section{Are galaxies supported by turbulent or thermal pressure?} \label{S:regimes}

We now use the scaling relations presented above to determine as a function of stellar mass
and redshift whether galaxies are supported by turbulent or thermal pressure (on average within
the effective radius).

\begin{figure*}
\centering
\includegraphics[width=0.95\columnwidth]{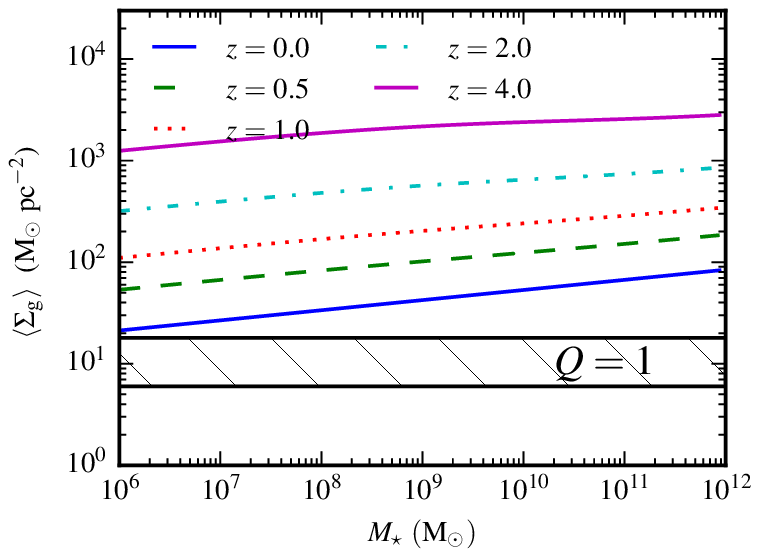}
\includegraphics[width=0.95\columnwidth]{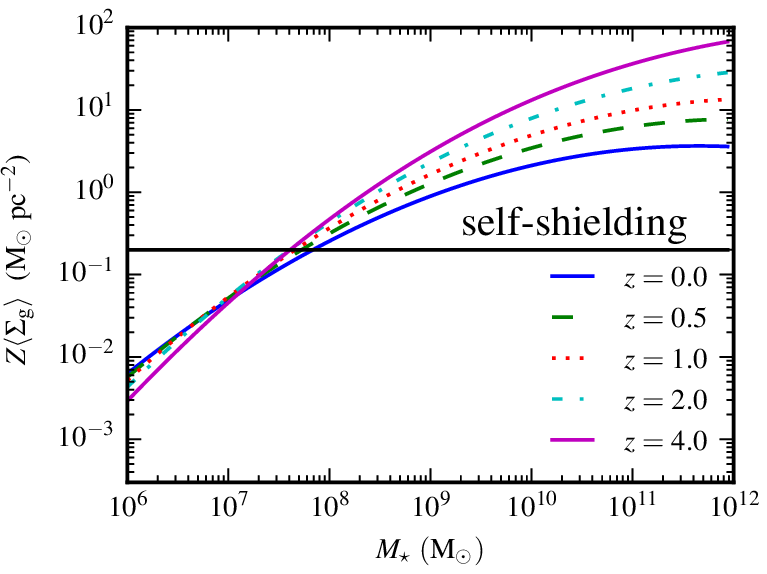}
\caption{Mean gas (\emph{left}) and metal (\emph{right}) surface density versus stellar mass at various redshifts calculated
using our empirically based scalings for $\fg$, $\re$, and $Z$. In the \emph{left} panel,
the hatched region indicates where the thermal Toomre $Q$ is unity (assuming $\Omega \sim 5-15$
Gyr$^{-1}$; see equation \ref{eq:q_eq_1}). Galaxies located above this region have sufficiently high gas surface densities that they cannot be
supported by thermal pressure alone. For the galaxy scaling relations that we assume, effectively all galaxies have $\qtherm < 1$
and thus cannot be globally in the thermal-pressure-dominated regime (although the outskirts of galaxies and galaxies
with lower-than-average $\msiggas$ values can be). In the \emph{right} panel, the black horizontal line indicates the
limit above which the galaxy is self-shielding. Independent of
redshift, the mean metal surface densities of $\mstar \ga 10^{8} \msun$ galaxies are greater than the self-shielding threshold.}
\label{fig:sig_gas_vs_mstar}
\end{figure*}

The gas surface density can be calculated using $\msiggas = \pi \re^2\fg/(1-\fg) \mstar$.
The left panel of \fref{fig:sig_gas_vs_mstar} shows $\msiggas$ versus $\mstar$ for various redshifts.
At fixed $z$, $\msiggas$ increases weakly with $\mstar$ (i.e. less than one order of magnitude over 8 orders of magnitude in stellar mass)
because the increase in $\msiggas$ that would result from the increase in total galaxy mass if all other quantities were held fixed is
almost completely mitigated by the decrease in $\fg$ and increase in $\re$ with $\mstar$. At fixed $\mstar$,
$\msiggas$ increases with $z$ because $\fg$ increases and $\re$ decreases. At $z = 0$, $\msiggas \sim 20-100 \msun$ pc$^{-2}$,
whereas at $z = 4$, $\msiggas \sim 1000-3000 \msun$ pc$^{-2}$.

To determine whether a galaxy is self-shielding, we require $Z \msiggas$, which can be calculated using the
scaling relations for $\fg$, $\re$, and $Z$. The right panel of \fref{fig:sig_gas_vs_mstar} shows the resulting relationship
between $Z\msiggas$ and $\mstar$ at various redshifts. The solid black horizontal line indicates
$Z\msiggas < 0.2 \msun$ pc$^{-2}$, the surface density above which the galaxy is self-shielding.
At fixed $z$, $Z \msiggas$ increases approximately linearly with $\mstar$ for $\mstar \la 10^8 \msun$, and the
relation flattens at higher stellar masses. At the low-mass end, $Z \msiggas$ increases decreases mildly with
increasing $z$ because the redshift evolution in $Z$ is stronger than that in $\msiggas$. At the high-mass end,
the opposite holds. For our present purposes, the most important feature to note is that independent
of redshift, galaxies with $\mstar \ga 10^8 \msun$ are self-shielding and thus globally in the turbulent-pressure-supported
regime.

In addition to not being self-shielding, a galaxy must have $Q > 1$ to be in the thermal-pressure-supported regime.
Given the values for $\Omega$ shown in the bottom-middle panel of \fref{fig:scaling_relations}, $Q > 1$ (see equation \ref{eq:q_eq_1}) corresponds to
$\msiggas \la 10 \msun$ pc$^{-2}$. From \fref{fig:sig_gas_vs_mstar}, we see that, at least assuming that the extrapolations
of our empirically based relations are reasonable, no galaxies have $\qtherm > 1$. Consequently, no galaxies in
equilibrium with typical properties should be globally thermal-pressure-supported. However, galaxies with lower-than-typical
$\siggas$ values can be in the thermal-pressure-supported regime, and it is possible that the outskirts of galaxies can
be in this regime even if the galaxy-averaged properties put it in the turbulent-pressure-supported regime.

\end{appendix}

\label{lastpage}

\end{document}